
\newcount\mgnf\newcount\tipi\newcount\tipoformule
\newcount\aux\newcount\driver\newcount\cind\global\newcount\bz
\newcount\tipobib\newcount\stile\newcount\modif
\newcount\noteno\noteno=1\newcount\figure

\newdimen\stdindent\newdimen\bibskip
\newdimen\maxit\maxit=0pt

\stile=0         
\tipobib=0       
\bz=0            
\cind=0          
\mgnf=0          
\tipoformule=0   
\aux=0           
\figure=1        


\ifnum\mgnf=0
   \magnification=\magstep0 
   \hsize=17truecm\vsize=24truecm\hoffset=-1cm
   \parindent=4.pt\stdindent=\parindent\fi
\ifnum\mgnf=1
   \magnification=\magstep1\hoffset=-0.85truecm
   \voffset=-1.0truecm\hsize=18.5truecm\vsize=24.truecm
   \baselineskip=14pt plus0.1pt minus0.1pt \parindent=6pt
   \lineskip=4pt\lineskiplimit=0.1pt      \parskip=0.1pt plus1pt
   \stdindent=\parindent\fi


\def\fine#1{}
\def\draft#1{\bz=1\ifnum\mgnf=1\baselineskip=22pt 
			      \else\baselineskip=16pt\fi
   \ifnum\stile=0\headline={\hfill DRAFT #1}\fi\raggedbottom
    \setbox150\vbox{\parindent=0pt\centerline{\bf Figures' captions}\*}
    \setbox149\vbox{\parindent=0pt\centerline{FIGURES}\*}
    \def\gnuins ##1 ##2 ##3{\gnuinsf {##1} {##2} {##3}}
    \def\gnuin ##1 ##2 ##3 ##4 ##5 ##6{\gnuinf {##1} {##2} {##3} 
		{##4} {##5} {##6}} 
    \def\eqfig##1##2##3##4##5##6{\eqfigf {##1} {##2} {##3} {##4} {##5}{##6}}
    \def\eqfigbis##1##2##3##4##5##6##7
	{\eqfigbisf {##1} {##2} {##3} {##4} {##5} {##6} {##7}}
    \def\eqfigfor##1##2##3##4##5##6##7
             {\eqfigforf {##1} {##2} {##3} {##4} {##5} {##6} {##7}}
      \def\fine ##1{\vfill\eject
	 	\def\geq(####1){}
               \unvbox150\vfill\eject\raggedbottom
                \unvbox149 ##1}}


\newcount\prau

\def\titolo#1{\setbox197\vbox{ 
\leftskip=0pt plus16em \rightskip=\leftskip
\spaceskip=.3333em \xspaceskip=.5em \parfillskip=0pt
\pretolerance=9999  \tolerance=9999
\hyphenpenalty=9999 \exhyphenpenalty=9999
\ftitolo #1}}
\def\abstract#1{\setbox198\vbox{
     \centerline{\vbox{\advance\hsize by -2cm \parindent=0pt\it Abstact: #1}}}}
\def\parole#1{\setbox195\hbox{
     \centerline{\vbox{\advance\hsize by -2cm \parindent=0pt Keywords: #1.}}}}
\def\autore#1#2{\setbox199\hbox{\unhbox199\ifnum\prau=0 #1%
\else, #1\fi\global\advance\prau by 1$^{\simbau}$}
     \setbox196\vbox {\parindent=0pt\copy196\item{$^{\simbau}$}{#2}}}
\def\prima{\unvbox197\vskip1truecm\centerline{\unhbox199}
     \footnote{}{\unvbox196}\vskip1truecm\unvbox198\vskip1truecm\copy195}
\def\simbau{\ifcase\prau
	\or \dagger \or \ddagger \or * \or \star \or \bullet\fi}


       \let\d=\delta 
      \let\l=\lambda
                   
\let\s=\sigma

\let\ge=\geq
\let\le=\leq


{\count255=\time\divide\count255 by 60 \xdef\oramin{\number\count255}
        \multiply\count255 by-60\advance\count255 by\time
   \xdef\oramin{\oramin:\ifnum\count255<10 0\fi\the\count255}}
\def\ora{\oramin }

\def\data{\number\day/\ifcase\month\or gennaio \or febbraio \or marzo \or
aprile \or maggio \or giugno \or luglio \or agosto \or settembre
\or ottobre \or novembre \or dicembre \fi/\number\year;\ \ora}

\setbox200\hbox{$\scriptscriptstyle \data $}


\newcount\pgn \pgn=1
\newcount\firstpage

\def\foglio{\number\numsec:\number\pgn\global\advance\pgn by 1}
\def\foglioa{A\number\numsec:\number\pgn\global\advance\pgn by 1}

\def\pagina{\vfill\eject}
\def\ppagina{\ifodd\pageno\pagina\null\pagina\else\pagina\fi}
\def\ppaginan{\ifodd-\pageno\pagina\null\pagina\else\pagina\fi}

\def\setind{\firstpage=\pageno}
\def\setcap#1{\null\def\titlecap{#1}\global\firstpage=\pageno}
\def\titletesi{Indici critici per sistemi fermionici in una dimensione}

\ifnum\stile=1
  \def\pagenumbers{\headline={%
  \ifnum\pageno=\firstpage\hfil\else%
     \ifodd\pageno\hfill{\sc\titlecap}~~{\bf\folio}%
      \else{\bf\folio}~~{\sc\titletesi}\hfill\fi\fi}
  \footline={\ifnum\bz=0
                   \hfill\else\rlap{\hbox{\copy200}\ $\st[\foglio]$}\hfill\fi}}
  \def\pagenumbersind{\headline={%
  \ifnum\pageno=\firstpage\hfil\else%
    \ifodd\pageno\hfill{\rm\romannumeral-\pageno}%
     \else{\rm\romannumeral-\pageno}\hfill\fi\fi}
  \footline={\ifnum\bz=0
                   \hfill\else\rlap{\hbox{\copy200}\ $\st[\foglio]$}\hfill\fi}}
\else
  \def\pagenumbers{\headline={\hfill}
     \footline={\ifnum\bz=0\hfill\folio\hfill
                \else\rlap{\hbox{\copy200}\ $\st[\foglio]$}
		   \hfill\folio\hfill\fi}}
\fi

\pagenumbers

\def\numeropag#1{
   \ifnum #1<0 \romannumeral -#1\else \number #1\fi
   }


\global\newcount\numsec\global\newcount\numfor
\global\newcount\numfig\global\newcount\numpar
\global\newcount\numteo\global\newcount\numlem

\numfig=1\numsec=0

\gdef\profonditastruttura{\dp\strutbox}
\def\senondefinito#1{\expandafter\ifx\csname #1\endcsname\relax}
\def\SIA #1,#2,#3 {\senondefinito{#1#2}%
\expandafter\xdef\csname#1#2\endcsname{#3}\else%
\write16{???? ma #1,#2 e' gia' stato definito !!!!}\fi}
\def\etichetta(#1){(\veroparagrafo.\veraformula)
\SIA e,#1,(\veroparagrafo.\veraformula)
 \global\advance\numfor by 1
\write15{\string\FU (#1){\equ(#1)}}
\9{ \write16{ EQ \equ(#1) == #1  }}}
\def \FU(#1)#2{\SIA fu,#1,#2 }
\def\etichettaa(#1){(A\veroparagrafo.\veraformula)
 \SIA e,#1,(A\veroparagrafo.\veraformula)
 \global\advance\numfor by 1
\write15{\string\FU (#1){\equ(#1)}}
\9{ \write16{ EQ \equ(#1) == #1  }}}
\def \FU(#1)#2{\SIA fu,#1,#2 }
\def\tetichetta(#1){\veroparagrafo.\veroteorema
\SIA e,#1,{\veroparagrafo.\veroteorema}
\global\advance\numteo by1
\write15{\string\FU (#1){\equ(#1)}}%
\9{\write16{ EQ \equ(#1) == #1}}}
\def\tetichettaa(#1){A\veroparagrafo.\veroteorema
\SIA e,#1,{A\veroparagrafo.\veroteorema}
\global\advance\numteo by1
\write15{\string\FU (#1){\equ(#1)}}%
\9{\write16{ EQ \equ(#1) == #1}}}
\def\letichetta(#1){\veroparagrafo.\verolemma
\SIA e,#1,{\veroparagrafo.\verolemma}
\global\advance\numlem by1
\write15{\string\FU (#1){\equ(#1)}}%
\9{\write16{ EQ \equ(#1) == #1}}}
\def\getichetta(#1){{\bf Fig. \verafigura}:
 \SIA e,#1,{\verafigura}
 \global\advance\numfig by 1
\write15{\string\FU (#1){\equ(#1)}}
\9{ \write16{ Fig. \equ(#1) ha simbolo  #1  }}}

\def\veroparagrafo{\number\numsec}\def\veraformula{\number\numfor}
\def\verafigura{\number\numfig}\def\veroteorema{\number\numteo}
\def\verolemma{\number\numlem}

\def\geq(#1){\getichetta(#1)\galato(#1)}
\def\Eq(#1){\eqno{\etichetta(#1)\alato(#1)}}
\def\eq(#1){&\etichetta(#1)\alato(#1)}
\def\Eqa(#1){\eqno{\etichettaa(#1)\alato(#1)}}
\def\eqa(#1){&\etichettaa(#1)\alato(#1)}
\def\teq(#1){\tetichetta(#1)\talato(#1)}
\def\teqa(#1){\tetichettaa(#1)\talato(#1)}
\def\leq(#1){\letichetta(#1)\talato(#1)}

\def\Eqr{\eqno(\veroparagrafo.\veraformula)\advance\numfor by 1}
\def\eqr{&(\veroparagrafo.\veraformula)\advance\numfor by 1}
\def\Eqar{\eqno(A\veroparagrafo.\veraformula)\advance\numfor by 1}
\def\eqar{&(A\veroparagrafo.\veraformula)\advance\numfor by 1}

\def\eqv(#1){\senondefinito{fu#1}$\clubsuit$$#1$\write16{Manca #1 !}%
\else\csname fu#1\endcsname\fi}
\def\equ(#1){\senondefinito{e#1}\eqv(#1)\else\csname e#1\endcsname\fi}


\newdimen\gwidth

\def\commenta#1{\ifnum\bz=1\strut \vadjust{\kern-\profonditastruttura
 \vtop to \profonditastruttura{\baselineskip
 \profonditastruttura\vss
 \rlap{\kern\hsize\kern0.1truecm
  \vbox{\hsize=1.7truecm\raggedright\nota\noindent #1}}}}\fi}
\def\talato(#1){\ifnum\bz=1\strut \vadjust{\kern-\profonditastruttura
 \vtop to \profonditastruttura{\baselineskip
 \profonditastruttura\vss
 \rlap{\kern-1.2truecm{$\scriptstyle#1$}}}}\fi}
\def\alato(#1){\ifnum\bz=1
 {\vtop to \profonditastruttura{\baselineskip
 \profonditastruttura\vss
 \rlap{\kern-\hsize\kern-1.2truecm{$\scriptstyle#1$}}}}\fi}
\def\galato(#1){\ifnum\bz=1 \gwidth=0pt 
 {\vtop to \profonditastruttura{\baselineskip
 \profonditastruttura\vss
 \rlap{\kern-\gwidth\kern-2.2truecm{$\scriptstyle#1$}}}}\fi}


\newskip\ttglue

\font\ftitolo=cmbx12 
\font\eighttt=cmtt8 \font\sevenit=cmti7  \font\sevensl=cmsl8
\font\sc=cmcsc10

\font\msytwww=msbm7 scaled\magstep1


\def\settepunti{\def\rm{\fam0\sevenrm}
\textfont0=\sevenrm \scriptfont0=\fiverm \scriptscriptfont0=\fiverm
\textfont1=\seveni \scriptfont1=\fivei   \scriptscriptfont1=\fivei
\textfont2=\sevensy \scriptfont2=\fivesy   \scriptscriptfont2=\fivesy
\textfont3=\tenex \scriptfont3=\tenex   \scriptscriptfont3=\tenex
\textfont\itfam=\sevenit  \def\it{\fam\itfam\sevenit}%
\textfont\slfam=\sevensl  \def\sl{\fam\slfam\sevensl}%
\textfont\ttfam=\eighttt  \def\tt{\fam\ttfam\eighttt}
\textfont\bffam=\sevenbf  \scriptfont\bffam=\fivebf
\scriptscriptfont\bffam=\fivebf  \def\bf{\fam\bffam\sevenbf}%
\tt \ttglue=.5em plus.25em minus.15em
\setbox\strutbox=\hbox{\vrule height6.5pt depth1.5pt width0pt}%
\normalbaselineskip=8pt\let\sc=\fiverm \normalbaselines\rm}

\let\nota=\settepunti

\def\nnn{\hbox{\msytwww N}}


\font\tenmib=cmmib10
\font\sevenmib=cmmib10 scaled 800

\textfont5=\tenmib  \scriptfont5=\sevenmib  \scriptscriptfont5=\fivei

\mathchardef\aaa= "050B
\mathchardef\xxx= "0518
\mathchardef\oo = "0521
\mathchardef\Dp = "0540
\mathchardef\H  = "0548
\mathchardef\FFF= "0546
\mathchardef\ppp= "0570
\mathchardef\nnn= "0517

\def\inputif#1{\ifnum \figure=1 \input #1\fi}
\newcount \caso

\newdimen\xshift \newdimen\xwidth \newdimen\yshift \newdimen\ywidth
\newdimen\laln

\def\ins#1#2#3{\nointerlineskip\vbox to0pt {\kern-#2 \hbox{\kern#1 #3}
\vss}}

\def\eqfig#1#2#3#4#5#6#7{
\xwidth=#1 \xshift=\hsize \advance\xshift 
by-\xwidth \divide\xshift by 2
\yshift=#2 \divide\yshift by 2
\caso=#7
\ifcase\caso\midinsert \or \topinsert \fi
\parindent=0pt
\line{\hglue\xshift \vbox to #2{\vfil 
#3 \includegraphics{#40.ps}
}\hfill}
\nobreak
\*
\didascalia{\geq(#6)#5}
\ifcase\caso\endinsert \or \endinsert \fi
}

\def\eqfigf#1#2#3#4#5#6#7{
\xwidth=#1 \xshift=\hsize \advance\xshift 
by-\xwidth \divide\xshift by 2
\yshift=#2 \divide\yshift by 2
\caso=#7
\ifcase\caso\midinsert \or \topinsert \fi
\parindent=0pt
\line{\hglue\xshift \vbox to #2{\vfil 
#3 \includegraphics{#40.ps}
}\hfill}
\nobreak
\*
\didascalia{\geq(#6)#5}
\ifcase\caso\endinsert \or \endinsert \fi
\setbox149\vbox{\unvbox149 \*\* \centerline{Fig. \equ(#6)} 
\nobreak
\*
\nobreak
\line{\hglue\xshift \vbox to #2{\vfil 
#3 \includegraphics{#40.ps}
}\hfill}\*}
\setbox150\vbox{\unvbox150 \parindent=0pt\*{\bf Fig. \equ(#6)}: #5\*}
}

\def\eqfigbisf#1#2#3#4#5#6#7{
\xwidth=#1 \multiply\xwidth by 2 
\xshift=\hsize \advance\xshift 
by-\xwidth \divide\xshift by 3
\divide\xwidth by 2 
\yshift=#2 
\ywidth=#2
\topinsert
\parindent=0pt
\hbox to \hsize{\hskip\xshift
\hbox to \xwidth{\vbox to \ywidth{\vfil#3\includegraphics{#50.ps}}\hfill}%
\hskip\xshift%
\hbox to \xwidth{\vbox to \ywidth{\vfil#4\includegraphics{#51.ps}}\hfill}\hfill}
\nobreak
\*\*
\didascalia{\geq(#7)#6}
\endinsert
\setbox149\vbox{\unvbox149 \*\* \centerline{Fig. \equ(#7)} 
\nobreak
\*
\nobreak
\parindent=0pt
\line{\hfill
\vbox to \ywidth{\vfil #3 \includegraphics{#50.ps}}
\hskip\xshift%
\vbox to \ywidth{\vfil #4 \includegraphics{#51.ps}}}
\*}
\setbox150\vbox{\unvbox150 \parindent=0pt\*{\bf Fig. \equ(#7)}: #6\*}
}

\def\eqfigbis#1#2#3#4#5#6#7{
\xwidth=#1 \multiply\xwidth by 2 
\xshift=\hsize \advance\xshift 
by-\xwidth \divide\xshift by 3
\divide\xwidth by 2 
\yshift=#2 
\ywidth=#2
\midinsert
\parindent=0pt
\hbox to \hsize{\hskip\xshift
\hbox to \xwidth{\vbox to \ywidth{\vfil#3\includegraphics{#50.ps}}\hfill}%
\hskip\xshift%
\hbox to \xwidth{\vbox to \ywidth{\vfil#4\includegraphics{#51.ps}}\hfill}\hfill}
\nobreak
\*\*
\didascalia{\geq(#7)#6}
\endinsert}

\def\dimenfor#1#2{\par\xwidth=#1 \multiply\xwidth by 2 
\xshift=\hsize \advance\xshift 
by-\xwidth \divide\xshift by 3
\divide\xwidth by 2 
\yshift=#2 
\ywidth=#2}

\def\eqfigfor#1#2#3#4#5#6#7{
\midinsert
\parindent=0pt
\hbox to \hsize{\hskip\xshift 
\hbox to \xwidth{\vbox to \ywidth{\vfil#1\includegraphics{#50.ps}}\hfill}%
\hskip\xshift%
\hbox to \xwidth{\vbox to \ywidth{\vfil#2\includegraphics{#51.ps}}\hfill}\hfill}
\nobreak
\line{\hglue\xshift 
\hbox to \xwidth{\vbox to \ywidth{\vfil #3 \includegraphics{#52.ps}}\hfill}%
\hglue\xshift
\hbox to \xwidth{\vbox to\ywidth {\vfil #4 \includegraphics{#53.ps}}\hfill}\hfill}
\nobreak
\*\*
\didascalia{\geq(#7)#6}
\endinsert}

\def\eqfigforf#1#2#3#4#5#6#7{
\midinsert
\parindent=0pt
\hbox to \hsize{\hskip\xshift 
\hbox to \xwidth{\vbox to \ywidth{\vfil#1\includegraphics{#50.ps}}\hfill}%
\hskip\xshift%
\hbox to \xwidth{\vbox to \ywidth{\vfil#2\includegraphics{#51.ps}}\hfill}\hfill}
\nobreak
\line{\hglue\xshift 
\hbox to \xwidth{\vbox to \ywidth{\vfil #3 \includegraphics{#52.ps}}\hfill}%
\hglue\xshift
\hbox to \xwidth{\vbox to\ywidth {\vfil #4 \includegraphics{#53.ps}}\hfill}\hfill}
\nobreak
\*\*
\didascalia{\geq(#7)#6}
\endinsert
\setbox149\vbox{\unvbox149\* \centerline{Fig. \equ(#7)} \nobreak\* \nobreak
\*
\vbox{\hbox to \hsize{\hskip\xshift 
\hbox to \xwidth{\vbox to \ywidth{\vfil#1\includegraphics{#50.ps}}\hfill}%
\hskip\xshift%
\hbox to \xwidth{\vbox to \ywidth{\vfil#2\includegraphics{#51.ps}}\hfill}\hfill}
\nobreak
\line{\hglue\xshift 
\hbox to \xwidth{\vbox to \ywidth{\vfil #3 \includegraphics{#52.ps}}\hfill}%
\hglue\xshift
\hbox to \xwidth{\vbox to\ywidth {\vfil #4 \includegraphics{#53.ps}}\hfill}\hfill}
}\hfill}
\setbox150\vbox{\unvbox150\parindent=0pt\* {\bf Fig.\equ(#7)}: #6\*}
}

\def\eqfigter#1#2#3#4#5#6#7{
\line{\hglue\xshift 
\vbox to \ywidth{\vfil #1 \includegraphics{#2.ps}}
\hglue30pt
\vbox to \ywidth{\vfil #3 \includegraphics{#4.ps}}\hfill}
\multiply\xshift by 3 \advance\xshift by \xwidth \divide\xshift by 2
\line{\hfill\hbox{#7}}
\line{\hglue\xshift 
\vbox to \ywidth{\vfil #5 \includegraphics{#6.ps}}}}


\def\7{\ifnum\modif=1\write13\else\write12\fi}
\def\8{\immediate\write13}


\def\gnuin #1 #2 #3 #4 #5 #6{\midinsert\vbox{\vbox to 260pt{
\hbox to 420pt{
\hbox to 200pt{\hfill\nota (a)\hfill}\hfill
\hbox to 200pt{\hfill\nota (b)\hfill}}
\vbox to 110pt{\vfill\hbox to 420pt{
\hbox to 200pt{\includegraphics{#1.ps}\hfill}\hfill
\hbox to 200pt{\includegraphics{#2.ps}\hfill}
}}\vfill
\hbox to 420pt{
\hbox to 200pt{\hfill\nota (c)\hfill}\hfill
\hbox to 200pt{\hfill\nota (d)\hfill}}
\vbox to 110pt{\vfill\hbox to 420pt{
\hbox to 200pt{\includegraphics{#3.ps}\hfill}\hfill
\hbox to 200pt{\includegraphics{#4.ps}\hfill}
}}\vfill}
\vskip0.25cm
\0\didascalia{\geq(#5)#6}}
\endinsert}

\def\gnuinf #1 #2 #3 #4 #5 #6{\midinsert\nointerlineskip\vbox to 260pt{
\hbox to 420pt{
\hbox to 200pt{\hfill\nota (a)\hfill}\hfill
\hbox to 200pt{\hfill\nota (b)\hfill}}
\vbox to 110pt{\vfill\hbox to 420pt{
\hbox to 200pt{\includegraphics{#1.ps}\hfill}\hfill
\hbox to 200pt{\includegraphics{#2.ps}\hfill}
}}\vfill
\hbox to 420pt{
\hbox to 200pt{\hfill\nota (c)\hfill}\hfill
\hbox to 200pt{\hfill\nota (d)\hfill}}
\vbox to 110pt{\vfill\hbox to 420pt{
\hbox to 200pt{\includegraphics{#3.ps}\hfill}\hfill
\hbox to 200pt{\includegraphics{#4.ps}\hfill}
}}\vfill}
\?
\0\didascalia{\geq(#5)#6}
\endinsert
\global\setbox150\vbox{\unvbox150 \*\*\0 Fig. \equ(#5): #6}
\global\setbox149\vbox{\unvbox149 \*\*
    \vbox{\centerline{Fig. \equ(#5)(a)} \nobreak
    \vbox to 200pt{\vfill\includegraphics{#1.ps_f}}}\*\*
    \vbox{\centerline{Fig. \equ(#5)(b)}\nobreak
    \vbox to 200pt{\vfill\includegraphics{#2.ps_f}}}\*\*
    \vbox{\centerline{Fig. \equ(#5)(c)}\nobreak
    \vbox to 200pt{\vfill\includegraphics{#3.ps_f}}}\*\*
    \vbox{\centerline{Fig. \equ(#5)(d)}\nobreak
    \vbox to 200pt{\vfill\includegraphics{#4.ps_f}}}
}}

\def\gnuins #1 #2 #3{\midinsert\nointerlineskip
\vbox{\line{\vbox to 220pt{\vfill
\includegraphics{#1.ps}}\hfill}
\*
\0\didascalia{\geq(#2)#3}}
\endinsert}

\def\gnuinsf #1 #2 #3{\midinsert\nointerlineskip
\vbox{\line{\vbox to 220pt{\vfill
\includegraphics{#1.ps}}\hfill}
\*
\0\didascalia{\geq(#2)#3}}\endinsert
\global\setbox150\vbox{\unvbox150 \*\*\0 Fig. \equ(#2): #3}
\global\setbox149\vbox{\unvbox149 \*\* 
    \vbox{\centerline{Fig. \equ(#2)}\nobreak
    \vbox to 220pt{\vfill\includegraphics{#1.ps_f}}}} 
}


\def\9#1{\ifnum\aux=1#1\else\relax\fi}
\let\numero=\number
\def\boh{\hbox{$\clubsuit$}\write16{Qualcosa di indefinito a pag. \the\pageno}}
\def\didascalia#1{\vbox{\nota\baselineskip=9truept\0#1\hfill}\vskip0.3truecm}
\def\frac#1#2{{#1\over #2}}
\def\V#1{\underline{#1}}

 	\let\0=\noindent
\def\guida{\leaders\hbox to 1em{\hss.\hss}\hfill}
\def\tende#1{\vtop{\ialign{##\crcr\rightarrowfill\crcr
              \noalign{\kern-1pt\nointerlineskip}
              \hglue3.pt${\scriptstyle #1}$\hglue3.pt\crcr}}}
\def\otto{{\kern-1.truept\leftarrow\kern-5.truept\to\kern-1.truept}}

\def\={{ \; \equiv \; }}		
\ifnum\mgnf=0
    \def\openone{\leavevmode\hbox{\ninerm 1\kern-3.3pt\tenrm1}}%
\fi
\ifnum\mgnf=1
     \def\openone{\leavevmode\hbox{\ninerm 1\kern-3.6pt\tenrm1}}%
\fi

\def\2{{1\over2}}

\def\igb{
    \mathop{\raise4.pt\hbox{\vrule height0.2pt depth0.2pt width6.pt}
    \kern0.3pt\kern-9pt\int}}

\def\st{\scriptscriptstyle}
\let\\=\noindent
\def\*{\vskip0.5truecm}
\def\?{\vskip0.75truecm}
\def\item#1{\vskip0.1truecm\parindent=0pt\par\setbox0=\hbox{#1\ }
     \hangindent\wd0\hangafter 1 #1 \parindent=\stdindent}

\def\annota#1#2{\footnote{${}^#1$}{\vtop {\hsize=\notesize\settepunti\baselineskip=8pt\parindent=0pt#2}}}
\def\annotano#1{\annota{\number\noteno}{#1}\advance\noteno by 1}

\newdimen\notesize\notesize=\hsize \advance \notesize by -\parindent


\def\ie{\hbox{\sl i.e.\ }}
\def\eg{\hbox{\sl e.g.\ }}
\def\qed{\hfill\break\nobreak\vbox{\vglue.25truecm\line{\hfill\raise1pt 
          \hbox{\vrule height9pt width5pt depth0pt}}}\vglue.25truecm}


\def\gint(#1)(#2)(#3){{\cal D}#1^{#2}\,e^{(#1^{#2+},#3#1^{#2-})}}

  \def\V0{{\bf 0}}    

\def\DD{{\cal E}}\def\EE{{\cal E}}
\def\GG{{\cal G}}\def\II{{\cal I}}
\def\JJ{{\cal J}}

\def\SS{{\cal S}}\def\TT{{\cal T}}


\ifnum\cind=1
\def\prtindex#1{\immediate\write\indiceout{\string\parte{#1}{\the\pageno}}}
\def\capindex#1#2{\immediate\write\indiceout{\string\capitolo{#1}{#2}{\the\pageno}}}
\def\parindex#1#2{\immediate\write\indiceout{\string\paragrafo{#1}{#2}{\the\pageno}}}
\def\subindex#1#2{\immediate\write\indiceout{\string\sparagrafo{#1}{#2}{\the\pageno}}}
\def\appindex#1#2{\immediate\write\indiceout{\string\appendice{#1}{#2}{\the\pageno}}}
\def\paraindex#1#2{\immediate\write\indiceout{\string\paragrafoapp{#1}{#2}{\the\pageno}}}
\def\subaindex#1#2{\immediate\write\indiceout{\string\sparagrafoapp{#1}{#2}{\the\pageno}}}
\def\bibindex#1{\immediate\write\indiceout{\string\bibliografia{#1}{Bibliografia}{\the\pageno}}}
\def\preindex#1{\immediate\write\indiceout{\string\premessa{#1}{\the\pageno}}}
\else
\def\prtindex#1{}
\def\capindex#1#2{}
\def\parindex#1#2{}
\def\subindex#1#2{}
\def\appindex#1#2{}
\def\paraindex#1#2{}
\def\subaindex#1#2{}
\def\bibindex#1{}
\def\preindex#1{}
\fi

\def\leaderfill{\leaders\hbox to 1em{\hss . \hss} \hfill }


\newdimen\capsalto \capsalto=0pt
\newdimen\parsalto \parsalto=20pt
\newdimen\sparsalto \sparsalto=30pt
\newdimen\tratitoloepagina \tratitoloepagina=2\parsalto
\def\aboveparteskip{\bigskip \bigskip}
\def\belowparteskip{\medskip \medskip}
\def\abovecapitskip{\bigskip}
\def\belowcapitskip{\medskip}
\def\belowparskip{\smallskip}
%


\def\parte#1#2{
   \9{\immediate\write16
      {#1     pag.\numeropag{#2} }}
   \aboveparteskip 
   \noindent 
   {\ftitolo #1} 
   \hfill {\ftitolo \numeropag{#2}}\par
   \belowparteskip
   }


\def\premessa#1#2{
   \9{\immediate\write16
      {#1     pag.\numeropag{#2} }}
   \abovecapitskip 
   \noindent 
   {\it #1} 
   \hfill {\rm \numeropag{#2}}\par
   \belowcapitskip
   }


\def\bibliografia#1#2#3{
  \ifnum\stile=1
   \9{\immediate\write16
      {Bibliografia    pag.\numeropag{#3} }}
   \belowcapitskip
   \noindent 
   {\bf Bibliografia} 
   \hfill {\bf \numeropag{#3}}\par
  \else
    \paragrafo{#1}{References}{#3}
\fi
   }


\newdimen\newstrutboxheight
\newstrutboxheight=\baselineskip
\advance\newstrutboxheight by -\dp\strutbox
\newdimen\newstrutboxdepth
\newstrutboxdepth=\dp\strutbox
\newbox\newstrutbox
\setbox\newstrutbox = \hbox{\vrule 
   height \newstrutboxheight 
   width 0pt 
   depth \newstrutboxdepth 
   }
\def\newstrut
   {\relax \ifmmode \copy \newstrutbox \else \unhcopy \newstrutbox \fi}
%
%
\vfuzz=3.5pt
%
%
\newdimen\indexsize \indexsize=\hsize
\advance \indexsize by -\tratitoloepagina
\newdimen\dummy
\newbox\parnum
\newbox\parbody
\newbox\parpage
%

%

\def\mastercap#1#2#3#4#5{
   \9{\immediate\write16
      {Cap. #3:#4     pag.\numeropag{#5} }}
   \abovecapitskip
   \setbox\parnum=\hbox {\kern#1\newstrut{#2}
			{\bf Capitolo~\number#3.}~}
   \dummy=\indexsize
   \advance\indexsize by -\wd\parnum
   \setbox\parbody=\vbox {
      \hsize = \indexsize \noindent \newstrut 
      {\bf #4}
      \newstrut \hss}
   \indexsize=\dummy
   \setbox\parnum=\vbox to \ht\parbody {
      \box\parnum
      \vfill 
      }
   \setbox\parpage = \hbox to \tratitoloepagina {
      \hss {\bf \numeropag{#5}}}
   \noindent \box\parnum\box\parbody\box\parpage\par
   \belowcapitskip
   }
\def\capitolo#1#2#3{\mastercap{\capsalto}{}{#1}{#2}{#3}}
%


%
\def\masterapp#1#2#3#4#5{
   \9{\immediate\write16
      {App. #3:#4     pag.\numeropag{#5} }}
   \abovecapitskip
   \setbox\parnum=\hbox {\kern#1\newstrut{#2}
			{\bf Appendice~A\number#3:}~}
   \dummy=\indexsize
   \advance\indexsize by -\wd\parnum
   \setbox\parbody=\vbox {
      \hsize = \indexsize \noindent \newstrut 
      {\bf #4}
      \newstrut \hss}
   \indexsize=\dummy
   \setbox\parnum=\vbox to \ht\parbody {
      \box\parnum
      \vfill 
      }
   \setbox\parpage = \hbox to \tratitoloepagina {
      \hss {\bf \numeropag{#5}}}
   \noindent \box\parnum\box\parbody\box\parpage\par
   \belowcapitskip
   }
\def\appendice#1#2#3{\masterapp{\capsalto}{}{#1}{#2}{#3}}
%

%

\def\masterpar#1#2#3#4#5{
   \9{\immediate\write16
      {par. #3:#4     pag.\numeropag{#5} }}
   \setbox\parnum=\hbox {\kern#1\newstrut{#2}\number#3.~}
   \dummy=\indexsize
   \advance\indexsize by -\wd\parnum
   \setbox\parbody=\vbox {
      \hsize = \indexsize \noindent \newstrut 
      #4
      \newstrut \hss}
   \indexsize=\dummy
   \setbox\parnum=\vbox to \ht\parbody {
      \box\parnum
      \vfill 
      }
   \setbox\parpage = \hbox to \tratitoloepagina {
      \hss \numeropag{#5}}
   \noindent \box\parnum\box\parbody\box\parpage\par
   \belowparskip
   }
\def\paragrafo#1#2#3{\masterpar{\parsalto}{}{#1}{#2}{#3}}
\def\sparagrafo#1#2#3{\masterpar{\sparsalto}{}{#1}{#2}{#3}}
%


%
\def\masterpara#1#2#3#4#5{
   \9{\immediate\write16
      {par. #3:#4     pag.\numeropag{#5} }}
   \setbox\parnum=\hbox {\kern#1\newstrut{#2}A\number#3.~}
   \dummy=\indexsize
   \advance\indexsize by -\wd\parnum
   \setbox\parbody=\vbox {
      \hsize = \indexsize \noindent \newstrut 
      #4
      \newstrut \hss}
   \indexsize=\dummy
   \setbox\parnum=\vbox to \ht\parbody {
      \box\parnum
      \vfill 
      }
   \setbox\parpage = \hbox to \tratitoloepagina {
      \hss \numeropag{#5}}
   \noindent \box\parnum\box\parbody\box\parpage\par
   \belowparskip
   }
\def\paragrafoapp#1#2#3{\masterpara{\parsalto}{}{#1}{#2}{#3}}
\def\sparagrafoapp#1#2#3{\masterpara{\sparsalto}{}{#1}{#2}{#3}}
%


\ifnum\stile=1

\def\newcap#1{\setcap{#1}
\vskip2.truecm\advance\numsec by 1
\\{\ftitolo \numero\numsec. #1}
\capindex{\numero\numsec}{#1}
\vskip1.truecm\numfor=1\pgn=1\numpar=1\numteo=1\numlem=1
}

\def\newapp#1{\setcap{#1}
\vskip2.truecm\advance\numsec by 1
\\{\ftitolo A\numero\numsec. #1}
\appindex{A\numero\numsec}{#1}
\vskip1.truecm\numfor=1\pgn=1\numpar=1\numteo=1\numlem=1
}

\def\newpar#1{
\vskip1.truecm
\vbox{
\\{\bf \numero\numsec.\numero\numpar. #1}
\parindex{\numero\numsec.\numero\numpar}{#1}
\*{}}
\nobreak
\advance\numpar by 1
}

\def\newpara#1{
\vskip1.truecm
\vbox{
\\{\bf A\numero\numsec.\numero\numpar. #1}
\paraindex{\numero\numsec.\numero\numpar}{#1}
\*{}}
\nobreak
\advance\numpar by 1
}

\else

\def\newsec#1{\vskip1.truecm
\advance\numsec by 1
\vbox{
\\{\bf \numero\numsec. #1}
\parindex{\numero\numsec}{#1}
\*{}}\numfor=1\pgn=1\numpar=1\numteo=1\numlem=1
\nobreak
}

\def\newsubsect#1{
\vskip1.truecm
\vbox{
\\{\bf \numero\numsec.\numero\numpar. #1}
\parindex{\numero\numsec.\numero\numpar}{#1}
\*{}}
\nobreak
\advance\numpar by 1
}

\def\newapp#1{\vskip1.truecm
\advance\numsec by 1
\vbox{
\\{\bf A\numero\numsec. #1}
\appindex{A\numero\numsec}{#1}
\*{}}\numfor=1\pgn=1\numpar=1\numteo=1\numlem=1
}

\def\biblio{\vskip1.truecm
\vbox{
\\{\bf References.}\*{}
\bibindex{{}}}\nobreak\makebiblio
}

\fi


\newread\indicein
\newwrite\indiceout

\def\faindice{
\openin\indicein=\jobname.ind
\ifeof\indicein\relax\else{
\ifnum\stile=1
  \pagenumbersind
  \pageno=-1
  \setind
  \null
  \vskip 2.truecm
  \\{\ftitolo Indice}
  \vskip 1.truecm
  \parskip = 0pt
  \input \jobname.ind
  \ppaginan
\else
\\{\bf Index}
\*{}
 \input \jobname.ind
\fi}\fi
\closein\indicein
\def\nomeindice{\jobname.ind}
\immediate\openout \indiceout = \nomeindice
}


\newwrite\bib
\immediate\openout\bib=\jobname.bib
\global\newcount\bibex
\bibex=0
\def\verabib{\number\bibex}

\ifnum\tipobib=0
\def\cita#1{\expandafter\ifx\csname c#1\endcsname\relax
\hbox{$\clubsuit$}#1\write16{Manca #1 !}%
\else\csname c#1\endcsname\fi}
\def\rife#1#2#3{\immediate\write\bib{\string\raf{#2}{#3}{#1}}
\immediate\write15{\string\C(#1){[#2]}}
\setbox199=\hbox{#2}\ifnum\maxit < \wd199 \maxit=\wd199\fi}
\else
\def\cita#1{%
\expandafter\ifx\csname d#1\endcsname\relax%
\expandafter\ifx\csname c#1\endcsname\relax%
\hbox{$\clubsuit$}#1\write16{Manca #1 !}%
\else\probib(ref. numero )(#1)%
\csname c#1\endcsname%
\fi\else\csname d#1\endcsname\fi}%
\def\rife#1#2#3{\immediate\write15{\string\Cp(#1){%
\string\immediate\string\write\string\bib{\string\string\string\raf%
{\string\verabib}{#3}{#1}}%
\string\Cn(#1){[\string\verabib]}%
\string\CCc(#1)%
}}}%
\fi

\def\Cn(#1)#2{\expandafter\xdef\csname d#1\endcsname{#2}}
\def\CCc(#1){\csname d#1\endcsname}
\def\probib(#1)(#2){\global\advance\bibex+1%
\9{\immediate\write16{#1\verabib => #2}}%
}

\def\C(#1)#2{\SIA c,#1,{#2}}
\def\Cp(#1)#2{\SIAnx c,#1,{#2}}

\def\SIAnx #1,#2,#3 {\senondefinito{#1#2}%
\expandafter\def\csname#1#2\endcsname{#3}\else%
\write16{???? ma #1,#2 e' gia' stato definito !!!!}\fi}

\bibskip=10truept
\def\hboxto{\hbox to}

\catcode`\{=12\catcode`\}=12
\catcode`\<=1\catcode`\>=2
\immediate\write\bib<
	\string\halign{\string\hboxto \string\maxit%
	{\string #\string\hfill}&%
        \string\vtop{\string\parindent=0pt\string\advance\string\hsize%
	by -1.55truecm%
	\string#\string\vskip \bibskip
	}\string\cr%
>
\catcode`\{=1\catcode`\}=2
\catcode`\<=12\catcode`\>=12

\def\raf#1#2#3{\ifnum \bz=0 [#1]&#2 \cr\else
\llap{${}_{\rm #3}$}[#1]&#2\cr\fi}

\newread\bibin

\catcode`\{=12\catcode`\}=12
\catcode`\<=1\catcode`\>=2
\def\chiudibib<
\catcode`\{=12\catcode`\}=12
\catcode`\<=1\catcode`\>=2
\immediate\write\bib<}>
\catcode`\{=1\catcode`\}=2
\catcode`\<=12\catcode`\>=12
>
\catcode`\{=1\catcode`\}=2
\catcode`\<=12\catcode`\>=12

\def\makebiblio{
\ifnum\tipobib=0
\advance \maxit by 10pt
\else
\maxit=1.truecm
\fi
\chiudibib
\immediate \closeout\bib
\openin\bibin=\jobname.bib
\ifeof\bibin\relax\else
\raggedbottom
\input \jobname.bib
\fi
}

\openin13=#1.aux \ifeof13 \relax \else
\input #1.aux \closein13\fi
\openin14=\jobname.aux \ifeof14 \relax \else
\input \jobname.aux \closein14 \fi
\immediate\openout15=\jobname.aux

\def\V#1{\underline{#1}}

\def\normalbaselines{\baselineskip=20pt\lineskip=3pt\lineskiplimit=3pt}

\def\bv{{\bf v}}
\def\bq{{\bf q}}
\def\bE{{\bf E}}

\def\bc{{\bf c}}
\def\bj{{\bf j}}


\titolo{Properties of Stationary Nonequilibrium States in the Thermostatted
Periodic Lorentz Gas I:  The One Particle System}

\autore{F. Bonetto}{Department of Mathematics, Rutgers University,
New Brunswick, NJ 08903; \hfill\break Present address: IHES, 35 Route de Chartres, 
91440 Bures-sur-Yvette, France, email:  {\tentt bonetto@ihes.fr}.}
\autore{D. Daems}{Department of Mathematics, Rutgers University,
New Brunswick, NJ 08903; \hfill\break 
Present address: Center for Nonlinear Phenomena and 
Complex Systems, Universit\'e Libre de Bruxelles, 1050 Bruxelles, Belgium, 
email: {\tentt ddaems@ulb.ac.be}.}
\autore{J.L. Lebowitz}{Department of Mathematics and Physics , 
Rutgers University, New Brunswick, NJ 08903, \hfill\break 
Present address: IHES, 35 Route de Chartres, 
91440 Bures-sur-Yvette, France, email: {\tentt lebowitz@ihes.fr}.}

\abstract{We study numerically and analytically the properties of the
stationary state of a particle moving under the influence of an
electric field $\bE$ in a two dimensional periodic Lorentz gas 
with the energy kept constant by a Gaussian
thermostat. Numerically the current appears to be a continuous
function of $\bE$ whose derivative varies very irregularly, possibly
in a discontinuous manner. We argue for the non differentibility of
the current as a function of $\bE$ utilizing a symbolic
description of the dynamics based on the discontinuities of the
collision map. The decay of correlations and the behavior of the
diffusion constant are also investigated.}

\parole{Thermostatted Lorentz gas, steady state current, smoothness,
regularity, symbolic dynamics}

\prima

\newsec{Introduction} 
There has been much interest in modeling stationary nonequilibrium
states (SNS) of physical systems by closed dynamical systems evolving
under a deterministic (time-reversible) non-Hamiltonian dynamics
\cita{H}\cita{EM}\cita{GC}.  This is in contrast to modeling by ``open'' 
systems with some stochastic dynamics at the boundaries or by
Hamiltonian coupling to infinite reservoirs
\cita{EPR1}\cita{EPR2}\cita{L}\cita{LS}.  Some such modeling is
necessary to obtain SNS since the evolution of an isolated system will
only permit equilibrium stationary states (or some very unstable
measures).

These dynamical system models modify the Hamiltonian time evolution by
the addition of purely formal (i.e. having no connection with the
actual dynamics of the system) thermostats such as a Gaussian
thermostat.  This keeps the kinetic or total energy of the
nonequilibrium system, e.g. one subjected to external forces which do
work on it and drive it away from equilibrium, constant in time
thereby permitting a SNS to exist.  Such a model of a stationary
current carrying state produced by a steady electric field $\bE$
acting on the charges in a conductor forming a circle (periodic
boundary conditions were first introduced by Moran and Hoover
\cita{MH}).  Various aspects of this model were later studied both
analytically and numerically \cita{R}\cita{CELS}.  It is however
still not clear how well this dynamical system with its nonphysical
way of extracting the heat generated by the current really models the
essential features of electrical conduction in a physical system.
This seems relevant to deciding on the utility of this
``thermostatted'' approach in nonequilibrium statistical mechanics.
To answer this question we present here and in subsequent works
detailed numerical and analytical studies of a class of such models
describing stationary states of current carrying
systems\cita{CELS}. The system studied in the present work is the
Moran-Hoover \cita{MH} model of a single particle moving among a
periodic two dimensional array of fixed scatterers (Sinai billiard) in
the presence of an external field $\bE$ and a Gaussian thermostat
which keeps its kinetic energy fixed, see Fig. \equ(Fig1).  In
\cita{BDLR},\cita{BDGL} we consider the generalization of this model to
$N$-particles, $N \ge 2$, see also \cita{BGG}.

\eqfig{300pt}{100pt}{
\ins{147pt}{12pt}{${\nota l}$}
\ins{148pt}{27pt}{${\nota \bv}$}
\ins{115pt}{38pt}{${\nota R_2}$}
\ins{140pt}{62pt}{${\nota R_1}$}
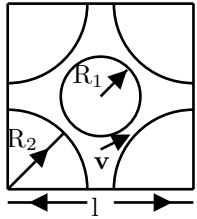
}{x1}{General billiard structure with scatterers of radius
$R_1$ and $R_2$ in a periodic box with side length $l$.
}{Fig1}0

The equations describing the motion of the particle, including
elastic scattering with the obstacles, are:

$$\left\{\eqalign{\dot\bq&=\bv\cr\dot\bv&=\bE-\alpha(\bv)\bv+F_{obs}(\bq)\cr
\alpha&={(\bv\cdot\bE)\over (\bv\cdot\bv)}}\right .\Eq(dyn)$$
where $F_{obs}(\bq)$ represents (symbolically) the collisions with the
obstacles, $(\;\cdot\;)$ is the particle equal to unity and ${\bf E} =
E \hat e$ with $|\hat e|=1$.

Eq.\equ(dyn) and thus the full dynamics keeps the kinetic energy
${\bv^2\over 2}$ fixed so that the motion takes place on the three
dimensional surface of constant energy which is the direct product of
a torus (with holes due to the obstacles) and  the circle with
$|\bv|=v_0$.  We can set $v=v_0\hat v$ where $\hat v$ is a unit vector
$\hat v=(cos\vartheta, \sin\vartheta)$. The microcanonical or uniform
distribution on this three dimensional energy surface is invariant
under the dynamics when $E=0$.  For $E\not=0$ the dynamics is not
Hamiltonian. There is in fact a phase space contraction $\gamma$ on
the energy surface equal to $(\bE\cdot\bv)/v_0^2$
\cita{MH}\cita{GG}\cita{CELS}.

Defining a new dimensionless time ${v_0\over l}t$ where $l$ is a
characteristic length of the system (say half the length of the torus) and
scaling $\bq$ by $l$, $\bv$ by $v_0$ and $E$ by ${l\over v_0^2}$ leave
the dynamics unchanged. We shall therefore take from now on $l=2$ and
$v_0=1$ (we choose $l=2$ because in this way we can see the obstacle
as having centers at $(0.0)$ and $(1,1)$). Then given any initial
absolutely continuous density $\psi_0(\vartheta,\bq)$ we shall be
interested in the time evolved density $\psi_t(\vartheta,\bq)$ in the
limit when $t\to\infty$ which describes the stationary nonequilibrium
state (SNS) of this system.

Moran and Hoover \cita{MH} carried out extensive numerical studies of
$\psi_t$, or more precisely of the induced measure $\nu_t$ on the
Poincar\'e cross section parametrized by the points
$P=(\alpha,\beta)\in S^1\times [-\pi/2,\pi/2]$, $\alpha$ 
corresponding to the angle locating
the point on the perimeter of the obstacle where the collision took
place and  $\beta$ to  the angle between the normal
vector at $\alpha$ and the unit velocity vector $\hat v$ (see subsection
2.1 for a more precise definition). Following the discrete time
trajectory $P_n$ (in their case there was only one obstacle per unit
cell of the triangular lattice, in the case of the billiard in
Fig. \equ(Fig1) one should add an integer component to $P_n$, \ie
$P_n=(\alpha,\beta,s)$ where $s\in\{0,1\}$ indicates which obstacle is
hit) they found that the density of points for large $n$ had a fractal
appearance, \ie the stationary measure appeared to be singular with
respect to the uniform (microcanonical) measure $(4 \pi)^{-1} \cos{\alpha}
 d\alpha d\phi$ on $P$ which is stationary
and approached (in a weak sense) as $t\to\infty$ when $\bE=0$.

This problem was later studied analytically by Chernov et
al. \cita{CELS} who proved that starting with any initial density
$\psi_0(\theta,{\bf q})$ there does indeed exist a limiting measure
$\nu_t\to\nu_E^+$ for the induced measure on the Poincar\'e section as
$t\to\infty$, whose Hausdorff dimension is, for $E\not=0$, strictly
less than two (the corresponding measure $\mu_E^+$ on the energy
surface has Hausdorff dimension less than three). They also showed that
the stationary current $\bj(\bf E)=\langle
\bv\rangle_{E}=\sigma_0\bE[1+o(|\bE|)]$ for $\bE$ small enough
$|{\bE}|<E_0$, $E_0 << 1$ (possibly as small as $10^{-20}$) where
$\langle \cdot\rangle_{E}$ is the average with respect to $\mu_E^+$
and $\sigma_0$ (generally a tensor) is given by the Kubo formula.  (It
also follows from their analysis that $\bj({\bE})$ is continuous for
$|{\bE}| < E_0$.) Recent work by Wojtkowski \cita{W} suggests that it
may be possible to extend the results of \cita{CELS} to larger value
of $|\bE|$; he found a precise value $E_0$ below wich the system is
hyperbolic and above which it is not.

In the present work we examine $\bj(\bE)$ and other properties of the
stationary measure $\nu^+$ for $E\in (0.025,2)$ and $\hat e=\hat
x$,\ie the field is in the $x$ direction and so $\bj({\bE})=j(E) {\hat x}$. 
Carrying out a numerical
evaluation of the Kubo formula for $\sigma_0$ (\ie an integral of the
autocorrelation function at zero field) we find good agreement with
the results of the simulation for $j(E)$ as $E$ tends towards
zero. A similar investigation, with less precision for small values of
$E$, was carried out for a triangular geometry in
\cita{R}, where the results, which look similar to ours for $E$ not
too small, were analyzed in terms of periodic orbits. Here we focus on
the behavior of the current for $E\in(0.025,0.5)$ where rather precise
numerical simulations indicate nonsmooth behavior of $j(E)$ vs. $E$.  In
particular we analyze the apparent singularities of $\nu^+_E$ as a
function of $E$ in terms of the change in the singularities of the map
from $P_n$ to $P_{n+1}$ caused by grazing collisions as $E$ changes.
We also analyze the formal expression for the derivatives of $j(E)$
obtained from the Kawasaki formula
\cita{CELS}, and the equivalent Ruelle formalism \cita{Ru}. We
identify possible origins of singularities and argue for a function
$j(E)$ that is non differentiable or, at most, admits a weak
derivative with a dense set of discontinuities although the non
rigorous nature of our argument does not permit a definitive statement
(a similar statement should hold for the dependence of $\bj(\bE)$ on $\hat e$).

It should be noted that, as shown in \cita{Po} the stationary current
is directly related to the sum of the stable and unstable Lyapunov
exponents of the system via the equality $(\bj(\bE)\cdot\bE)=
-[\lambda^s (\bE) +\lambda^u (\bE)]$.  There is also a relation
between the Hausdorff dimension of $\mu^+$ and the Lyapunov exponents
$D_H(\nu^+) = 1-\lambda^u (\bE)/\lambda^s(\bE)$ (remember that
$\lambda^s(0)+\lambda^u(0)=0$ but $\lambda^s(\bE)+\lambda^u(\bE)<0$ for
$E\not=0$).

In addition to $\bj(\bE)$ we also investigate the diffusion constant $d(E)$
relative to the drift as a function of $E$ and compare it to
$d\bj(E)/dE$, equality constituting in the limit $E\to0$ the Einstein
relation proven in
\cita{CELS}. Finally we compute the decay of the velocity
autocorrelation function for $E\not=0$ and analyze the numerical
results using the methods developed in \cita{GG}. We find that the
decay continues to be exponential (this was proven for $E=0$ in
\cita{Y}) with an oscillatory behavior which changes qualitatively at
or close to those values of $E$ at which we see the nonsmooth behavior
in the higher derivatives of $\bj(\bE)$.

\newsec{Current versus field: results and discussion}

We first describe the results of our simulations and then discuss possible
theoretical explanations of their most interesting features.

\newsubsect{Numerical results}

We used three different methods to compute the current as a function
of the field for the model in Fig. \equ(Fig1) with $R_1=0.39$,
$R_2=0.79$\annotano{These values coincide with the one used in
\cita{BGG}. There a system with many particles was studied and the
radia were set to $R_1=0.4$, $R_2=0.8$ plus a particles' radius of
$r=0.01$. It is clear that if only one particle is present this is
equivalent to a point particle moving among scatterers with radii
$R_1=0.39$ and $R_2=0.79$.}\ and $|\bv|=1$. The first two are based on the
fact that the system appears to be ergodic for
$E^{<}\hskip-1.5ex_{\sim} 2$ (this has been proved only for the case
when the field is very small, see \cita{CELS}). This allows us to
compute the current by choosing a ``typical'' initial point and
evolving it for a very long time while measuring the time average of
the instantaneous current. 

\dimen2=25pt \advance\dimen2 by 269pt
\dimen3=398pt
\eqfig{\dimen3}{314pt}{
\ins{378pt}{0pt}{$\hbox to20pt{\hfill ${\scriptstyle E}$\hfill}$}
\ins{24pt}{\dimen2}{$\hbox to25pt{\hfill ${\scriptstyle \kappa(E)}$\hfill}$}
\ins{194pt}{287pt}{$\hbox to174pt{\hfill ${\scriptstyle \kappa(E)}$\hfill}$}
\ins{65pt}{3pt}{$\hbox to 39pt{\hfill${\scriptstyle 0.025}$\hfill}$}
\ins{105pt}{3pt}{$\hbox to 39pt{\hfill${\scriptstyle 0.050}$\hfill}$}
\ins{144pt}{3pt}{$\hbox to 39pt{\hfill${\scriptstyle 0.075}$\hfill}$}
\ins{184pt}{3pt}{$\hbox to 39pt{\hfill${\scriptstyle 0.100}$\hfill}$}
\ins{224pt}{3pt}{$\hbox to 39pt{\hfill${\scriptstyle 0.125}$\hfill}$}
\ins{263pt}{3pt}{$\hbox to 39pt{\hfill${\scriptstyle 0.150}$\hfill}$}
\ins{303pt}{3pt}{$\hbox to 39pt{\hfill${\scriptstyle 0.175}$\hfill}$}
\ins{342pt}{3pt}{$\hbox to 39pt{\hfill${\scriptstyle 0.200}$\hfill}$}
\ins{0pt}{45pt}{$\hbox to 26pt{\hfill${\scriptstyle 0.170}$}$}
\ins{0pt}{69pt}{$\hbox to 26pt{\hfill${\scriptstyle 0.171}$}$}
\ins{0pt}{92pt}{$\hbox to 26pt{\hfill${\scriptstyle 0.172}$}$}
\ins{0pt}{115pt}{$\hbox to 26pt{\hfill${\scriptstyle 0.173}$}$}
\ins{0pt}{139pt}{$\hbox to 26pt{\hfill${\scriptstyle 0.174}$}$}
\ins{0pt}{162pt}{$\hbox to 26pt{\hfill${\scriptstyle 0.175}$}$}
\ins{0pt}{185pt}{$\hbox to 26pt{\hfill${\scriptstyle 0.176}$}$}
\ins{0pt}{209pt}{$\hbox to 26pt{\hfill${\scriptstyle 0.177}$}$}
\ins{0pt}{232pt}{$\hbox to 26pt{\hfill${\scriptstyle 0.178}$}$}
\ins{0pt}{255pt}{$\hbox to 26pt{\hfill${\scriptstyle 0.179}$}$}
\ins{0pt}{279pt}{$\hbox to 26pt{\hfill${\scriptstyle 0.180}$}$}
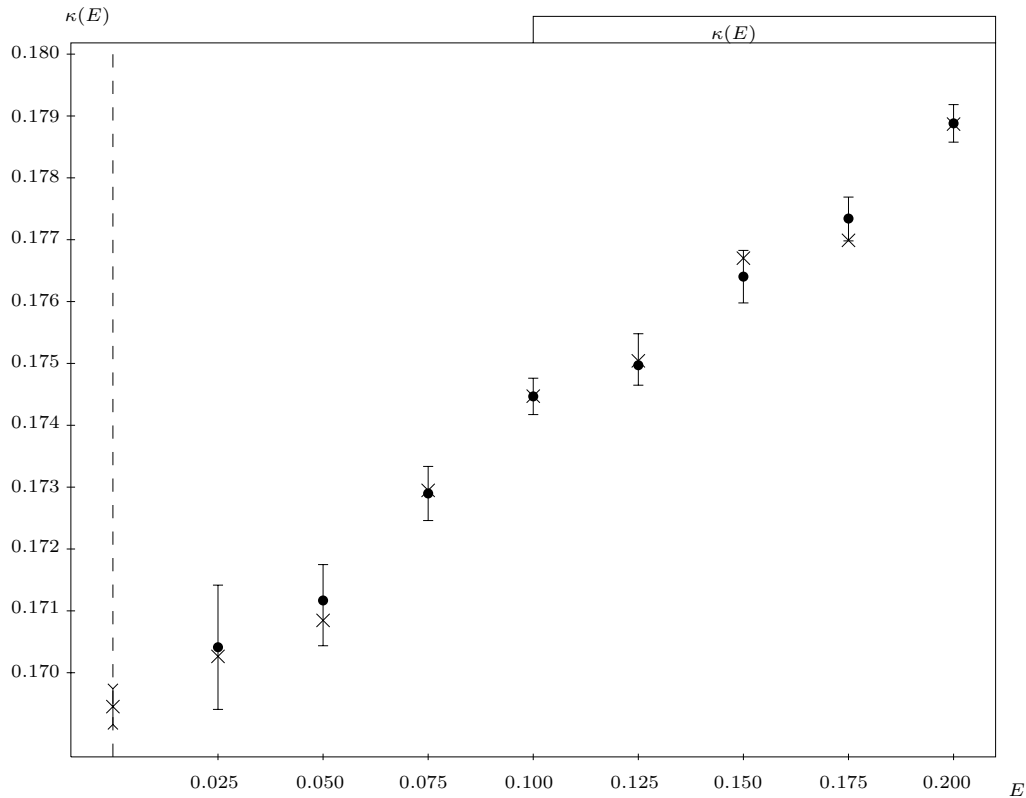
}{corrente_c}{Conductivity as a function of the field $E$. The filled circles 
represent the results from the time average on a single long trajectory with 
their error-bar. The crosses represent the result from the 
Kawasaki simulation. No error-bar is reported in this case (except for the 
value at $E=0$) to enhance readability. In any case the errors on 
the Kawasaki values are larger than the ones on the long time averages.}
{fig_cor}0

The simulations were done by computing a trajectory of $10^9$ successive
collisions. The only difficulty in the algorithm is in computing the
successive collision times in a fast enough manner without missing
collisions that are nearly tangent. Our algorithm is based on a Newton
method: it takes around 20 hours to compute $10^9$ successive
collisions on a i586 processor with a 350 Mhz clock under a Linux
operating system.  The values of $\kappa(E)=j(E)/E$ obtained by the
simulation are represented by the filled points in Fig. \equ(fig_cor).
The error-bars are computed by running 10 independent initial
conditions and looking at the maximum and minimum values obtained in
this way. It is reasonable to believe that, while the value of the
current averaged over a long time $T$ goes to 0 as the field goes to
0, the fluctuations about that average with respect to the initial
point should be more or less independent of $E$. Considering the heavy
numerical work needed to evolve 10 points for $10^9$ successive
collisions we did it for $E=0.025$, $0.05$, $0.1$, $0.125$ and $0.175$
and interpolated linearly for the error-bar at other points (the
reason of the choice to leave out only $E=0.75$, $0.150$ and $0.2$ is
because we think nothing new happens there in a sense that will become
clear in the following subsection).

To obtain more confidence in our numerics we also computed $\kappa(E)$
using the Kawasaki formula which allows us to compute also $\kappa(0)$
(for a rigorous proof of this formula when $E$ is very small see
\cita{CELS}).
Letting $S_E(t,X)$, with $X=(\bq,\bv)$, be the evolution generated by
\equ(dyn) and calling $J(X)$ the velocity component along the
direction of $\bE$,  then $j(E)=\langle J(X) \rangle_E$ with $\langle
\cdot \rangle_E$ representing the average with respect to $\mu_E^+$,
one has the Kawasaki formula\cita{CELS},

$$\kappa(E)=
\int_{0}^\infty dt \langle J(X)J(S_E(t,X)) \rangle_0\Eq(kaw)$$
where $\langle\cdot\rangle_0$ represents the average with respect to
the microcanonical distribution corresponding to $\mu^+_0$. Observe
that eq.\equ(kaw) reduces for $E=0$ to the well known Green-Kubo
expression for the linear response to an external field.

There are two problems in using eq.\equ(kaw) numerically. The first is
the necessity of a very good random number generator for the initial
conditions and the second is the impossibility of integrating the
correlation function for long times if one wants to have good
statistics. We solved the first problem by using a R250 generator (see
\cita{NR}) and, to ensure a better independence, we decorrelated the
initial points by evolving then for 40 collisions without field. In
this situation we observed that at a time equal to 50 times the mean
free flight time the correlation function in eq.\equ(kaw) is of the
order of $10^{-4}$ allowing us to say that the truncation error is
probably much smaller than the statistical error due to the fact that
we evolved only $2\cdot 10^6$ initial conditions. Observe that in this
case we can assume that our experiment consists in the independent
sampling of a random variable so that we can estimate the statistical
error by its standard deviation divided by the square root of the
number of initial points.  We observe finally that this method,
although numerically demanding, is theoretically more reliable than
the first method used to get the filled points in Fig. \equ(fig_cor)
due to the intrinsic instability of the dynamics under consideration,
see
\cita{GG} for a more precise discussion.

The results of this simulation are also plotted in Fig. \equ(fig_cor)
using crosses. We report the error bar only for the value at $E=0$ to
maintain readability of the graphs. We note that the error-bars for
$E>0$ are approximatively twice that at $E=0$ (simulations at
$E=0$ are much easier and faster).  This means that all the values are
in within the error bar of those computed with the previous method.

Due to the symmetry of our system we clearly have that
$\kappa(E)=\kappa(-E)$ so, assuming that $\kappa(E)$ is a smooth
function of $E$ we should be able to fit it for small $E$ by
$\kappa(E)=\kappa(0)+\kappa''(0)E^2$. Such a fit is indeed possible
(in particular one would find $\kappa(0)=0.169$ and
$\kappa''(0)=0.53$) but it will pass only through the error-bars of
the first three points. From the fourth point on the graph has a very
linear appearance (to be precise we can fit it well by
$\kappa(E)=0.169+0.042E$). The problem with this linear fit is that it
would produce a discontinuity of the first derivative of $\kappa(E)$
at some value of $E\in(0.05,0.075)$. Moreover, due to the very large
error-bars near $E=0$, it is easy to see that the linear fit
for $E>0.075$ passes through the error-bars of all the points leaving
open the possibility of a discontinuity of the first derivative of
$j(E)$ at $E=0$.

Clearly the question of discontinuities in the derivatives  
cannot be decided on the basis of numerical simulations. We need 
an analytical argument (to be transformed hopefully into a proof) to
decide this question. We present such an argument in
the following subsection. But we first show in Fig. \equ(figh) a graph of 
the current for higher value of the field to see if other
discontinuities of the derivative can be seen. We used the first method 
described before Fig. \equ(fig_cor) to compute the current from $E=0.2$ to
$E=2.0$ at steps of $\Delta E=0.025$. Each run consisted of $5\cdot
10^7$ collisions and the error bars have been computed for values of 
$E=0.25+0.25i$ by running 10 initial condition and interpolated
for the other points (when they are not visible it is because they are
smaller than the points).  

\dimen2=25pt \advance\dimen2 by 269pt
\dimen3=398pt
\eqfig{\dimen3}{314pt}{
\ins{378pt}{0pt}{$\hbox to20pt{\hfill ${\scriptstyle E}$\hfill}$}
\ins{24pt}{\dimen2}{$\hbox to25pt{\hfill ${\scriptstyle \kappa(E)}$\hfill}$}
\ins{194pt}{287pt}{$\hbox to174pt{\hfill ${\scriptstyle \kappa(E)}$\hfill}$}
\ins{37pt}{3pt}{$\hbox to 17pt{\hfill${\scriptstyle 0.2}$\hfill}$}
\ins{54pt}{3pt}{$\hbox to 17pt{\hfill${\scriptstyle 0.3}$\hfill}$}
\ins{72pt}{3pt}{$\hbox to 17pt{\hfill${\scriptstyle 0.4}$\hfill}$}
\ins{89pt}{3pt}{$\hbox to 17pt{\hfill${\scriptstyle 0.5}$\hfill}$}
\ins{107pt}{3pt}{$\hbox to 17pt{\hfill${\scriptstyle 0.6}$\hfill}$}
\ins{125pt}{3pt}{$\hbox to 17pt{\hfill${\scriptstyle 0.7}$\hfill}$}
\ins{142pt}{3pt}{$\hbox to 17pt{\hfill${\scriptstyle 0.8}$\hfill}$}
\ins{160pt}{3pt}{$\hbox to 17pt{\hfill${\scriptstyle 0.9}$\hfill}$}
\ins{177pt}{3pt}{$\hbox to 17pt{\hfill${\scriptstyle 1.0}$\hfill}$}
\ins{195pt}{3pt}{$\hbox to 17pt{\hfill${\scriptstyle 1.1}$\hfill}$}
\ins{213pt}{3pt}{$\hbox to 17pt{\hfill${\scriptstyle 1.2}$\hfill}$}
\ins{230pt}{3pt}{$\hbox to 17pt{\hfill${\scriptstyle 1.3}$\hfill}$}
\ins{248pt}{3pt}{$\hbox to 17pt{\hfill${\scriptstyle 1.4}$\hfill}$}
\ins{265pt}{3pt}{$\hbox to 17pt{\hfill${\scriptstyle 1.5}$\hfill}$}
\ins{283pt}{3pt}{$\hbox to 17pt{\hfill${\scriptstyle 1.6}$\hfill}$}
\ins{301pt}{3pt}{$\hbox to 17pt{\hfill${\scriptstyle 1.7}$\hfill}$}
\ins{318pt}{3pt}{$\hbox to 17pt{\hfill${\scriptstyle 1.8}$\hfill}$}
\ins{336pt}{3pt}{$\hbox to 17pt{\hfill${\scriptstyle 1.9}$\hfill}$}
\ins{353pt}{3pt}{$\hbox to 17pt{\hfill${\scriptstyle 2.0}$\hfill}$}
\ins{0pt}{31pt}{$\hbox to 26pt{\hfill${\scriptstyle 0.120}$}$}
\ins{0pt}{50pt}{$\hbox to 26pt{\hfill${\scriptstyle 0.125}$}$}
\ins{0pt}{70pt}{$\hbox to 26pt{\hfill${\scriptstyle 0.130}$}$}
\ins{0pt}{89pt}{$\hbox to 26pt{\hfill${\scriptstyle 0.135}$}$}
\ins{0pt}{109pt}{$\hbox to 26pt{\hfill${\scriptstyle 0.140}$}$}
\ins{0pt}{128pt}{$\hbox to 26pt{\hfill${\scriptstyle 0.145}$}$}
\ins{0pt}{148pt}{$\hbox to 26pt{\hfill${\scriptstyle 0.150}$}$}
\ins{0pt}{167pt}{$\hbox to 26pt{\hfill${\scriptstyle 0.155}$}$}
\ins{0pt}{187pt}{$\hbox to 26pt{\hfill${\scriptstyle 0.160}$}$}
\ins{0pt}{206pt}{$\hbox to 26pt{\hfill${\scriptstyle 0.165}$}$}
\ins{0pt}{226pt}{$\hbox to 26pt{\hfill${\scriptstyle 0.170}$}$}
\ins{0pt}{245pt}{$\hbox to 26pt{\hfill${\scriptstyle 0.175}$}$}
\ins{0pt}{265pt}{$\hbox to 26pt{\hfill${\scriptstyle 0.180}$}$}
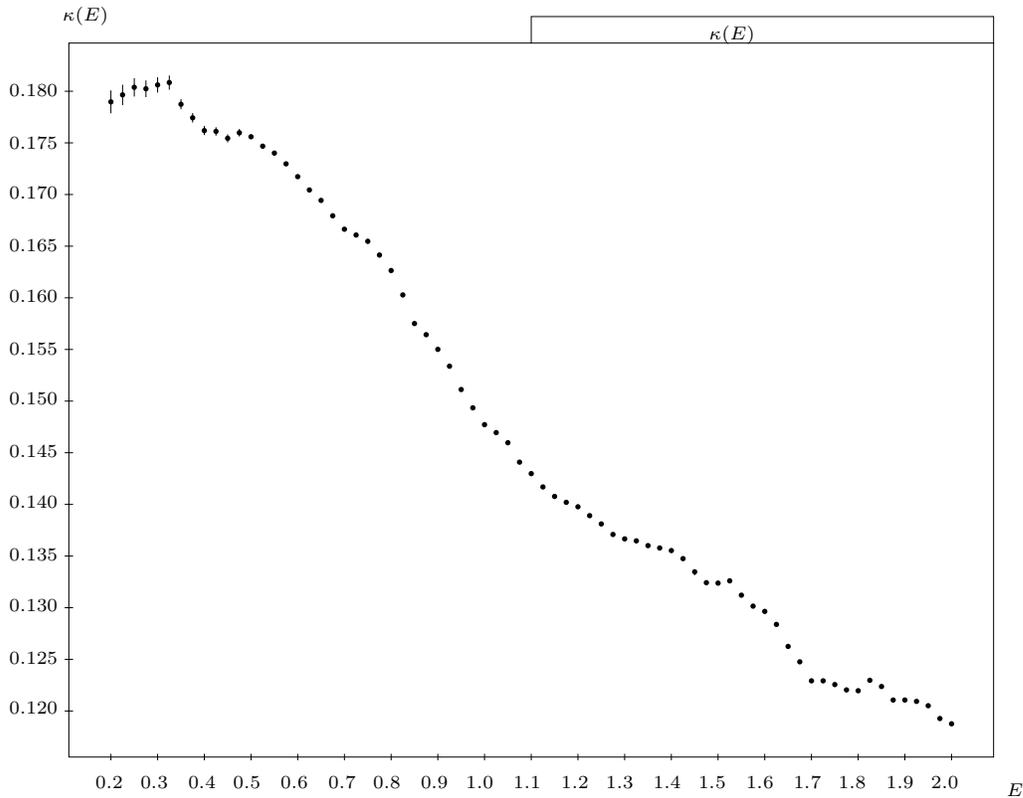
}{corrente_h}{Behavior of the current for higher value of the field. 
When no error-bar is visible it means that it is smaller than the symbol 
used.}{figh}0

Based on Fig. \equ(figh) it seems consistent to assume that $\kappa(E)$
is continuous for $E<2$. Its derivative on the contrary continues to
change in a very irregular, possibly discontinuous, way. The first
value of $E$ at which a big change is seen is around $E=0.35$. To
better appreciate this we plot the conductivity for $E=0.2$ to $E=0.5$
in Fig. \equ(figcd).
%
\dimen2=25pt \advance\dimen2 by 269pt
\dimen3=398pt
\eqfig{\dimen3}{314pt}{
\ins{378pt}{0pt}{$\hbox to20pt{\hfill ${\scriptstyle E}$\hfill}$}
\ins{24pt}{\dimen2}{$\hbox to25pt{\hfill ${\scriptstyle \kappa(E)}$\hfill}$}
\ins{194pt}{287pt}{$\hbox to174pt{\hfill ${\scriptstyle \kappa(E)}$\hfill}$}
\ins{19pt}{3pt}{$\hbox to 52pt{\hfill${\scriptstyle 0.20}$\hfill}$}
\ins{72pt}{3pt}{$\hbox to 52pt{\hfill${\scriptstyle 0.25}$\hfill}$}
\ins{125pt}{3pt}{$\hbox to 52pt{\hfill${\scriptstyle 0.30}$\hfill}$}
\ins{177pt}{3pt}{$\hbox to 52pt{\hfill${\scriptstyle 0.35}$\hfill}$}
\ins{230pt}{3pt}{$\hbox to 52pt{\hfill${\scriptstyle 0.40}$\hfill}$}
\ins{283pt}{3pt}{$\hbox to 52pt{\hfill${\scriptstyle 0.45}$\hfill}$}
\ins{336pt}{3pt}{$\hbox to 52pt{\hfill${\scriptstyle 0.50}$\hfill}$}
\ins{0pt}{24pt}{$\hbox to 26pt{\hfill${\scriptstyle 0.175}$}$}
\ins{0pt}{62pt}{$\hbox to 26pt{\hfill${\scriptstyle 0.176}$}$}
\ins{0pt}{100pt}{$\hbox to 26pt{\hfill${\scriptstyle 0.177}$}$}
\ins{0pt}{137pt}{$\hbox to 26pt{\hfill${\scriptstyle 0.178}$}$}
\ins{0pt}{175pt}{$\hbox to 26pt{\hfill${\scriptstyle 0.179}$}$}
\ins{0pt}{213pt}{$\hbox to 26pt{\hfill${\scriptstyle 0.180}$}$}
\ins{0pt}{251pt}{$\hbox to 26pt{\hfill${\scriptstyle 0.181}$}$}
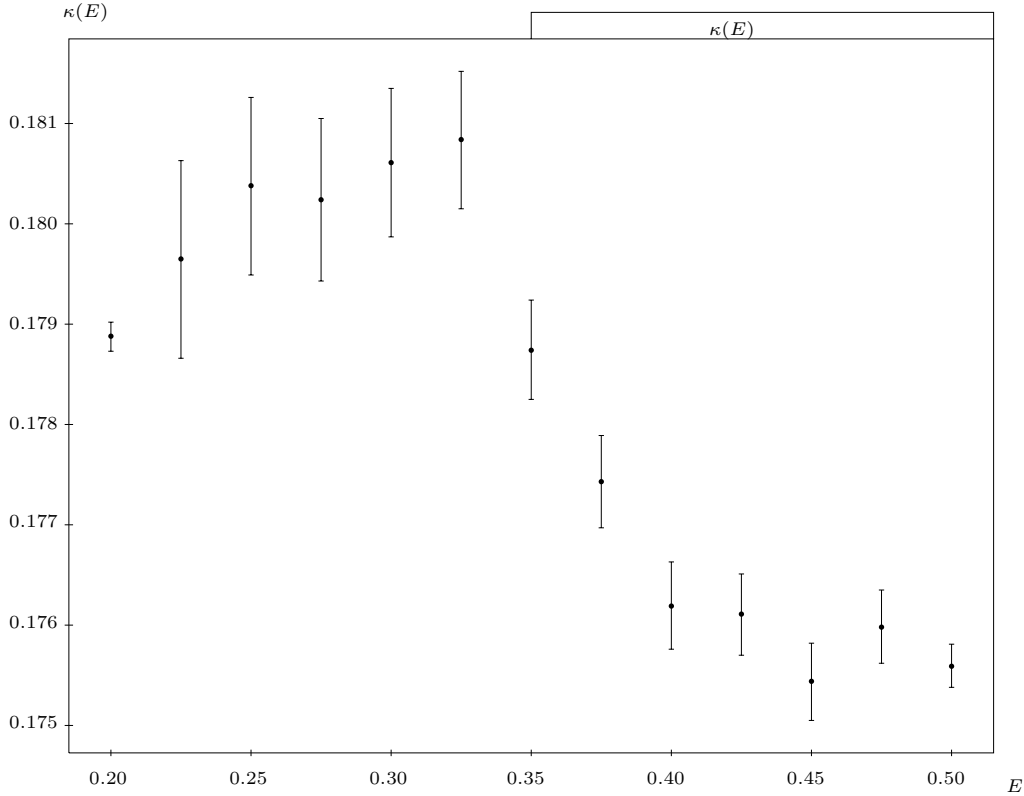
}{corrente_d}{Current versus field for $E\in[0.2,0.5]$. The intervals 
$(0.225,0.25)$ and $(0.325, 0.35)$ and $(0.375, 0.425)$ possibly contain 
discountinuity of the first derivative.}{figcd}0
As can be easily seen the most probable site of discontinuous behavior
of the derivative are at fields in the intervals $(0.225, 0.25)$,
$(0.325, 0.35)$ and $(0.375,0.425)$. After discussing the possible
origin of non smooth behavior of the expectation of a smooth function with
respect to $\mu^+_E$ we will return to the behavior of $j(E)$ at those
values of $E$. 

We briefly report on a third method to compute the value of the
conductivity that we used to be sure that the above results are not
due to a bias in our Newton scheme. The idea is the same as the first
method but instead of the fast Newton style algorithm we used a
discrete time integrator (a fourth order Runge-Kutta integrator). To
avoid missing collisions that are almost tangent we had to choose a
very small time step, \ie $\delta t=10^{-4}$. The slowness of this
algorithm permitted us to check only a few points and they were
consistent with the other methods.

We close this section observing that our simulations support the
validity of a central limit theorem for our system. In fact if one
looks at the behavior of the maximum and minimum values observed among
the ten runs as a function of the running time $T$ one finds that one
can fit it with a function of the form $C/\sqrt{T}$ with
$C^2\simeq\langle J(X)^2\rangle_E-\langle J(X)\rangle^2_E$.

\newsubsect{Analytical discussion}

To study the properties of $\kappa(E)$ as a function of $E$ we can use
eq.\equ(kaw). We know that the correlation functions in eq.\equ(kaw)
are uniformly integrable for $E<E_0$ for some very small $E_0$, see
\cita{CELS}. This allows us to exchange limits,

$$\eqalign{\lim_{E\to E'}\int_{0}^\infty dt \langle &J(X)J(S_E(t,X)) 
\rangle_0=\cr
\int_{0}^\infty dt &\langle J(X)\lim_{E\to E'}J(S_E(t,X)) \rangle_0=\cr
&\int_{0}^\infty dt \langle J(X)J(S_{E'}(t,X)) \rangle_0}\Eq(cont)$$
and conclude that $\kappa(E)$ is a continuous function of $E$ for
$E<E_0$. Although the argument in \cita{CELS} gives, as already stated,
a very small value for $E_0$ the numerical results, some of which are
reported at the end of this subsection, support the validity of the
above reasoning at least for $E^{<}\hskip-1.5ex_{\sim} 1$.

Differentiating eq.\equ(kaw) to obtain the higher derivatives of
$\kappa(E)$ as a function of $E$ is essentially equivalent to what is
formally done in \cita{Ru}. We obtain
$${d\kappa(E)\over dE}=
\int_0^\infty dt\int_0^t dt' 
\langle J(X)F(S_E(t',X))\cdot
\nabla_{S_E(t',X)}J(S_E(t-t',S_E(t',X))) \rangle_0.
\Eq(dkaw0)$$
where $F(X)=\hat x-(\hat x\cdot\bv)\bv$ is the derivative respect to
$E$ of the right hand side of eq.\equ(dyn). This can be rewritten as:

$$\eqalign{{d\kappa(E)\over dE}=&
\int_0^\infty dt\int_0^t dt'
\langle J(X)J(S_E(t',X))J(S_E(t,X))) \rangle_0\cr
+&\int_0^\infty dt\int_0^t dt'
\langle\nabla_{S_E(t',X)}J(X)\cdot F(S_E(t',X))J(S_E(t,X))) 
\rangle_0\cr
}\Eq(dkaw1)$$
Observe that the second term on the right hand side can be rewritten as:
$$\eqalign{\int_0^\infty dt\int_0^t dt'
\langle\nabla_{X}J(X)&\Lambda^{-1}(t',X)F(S_E(t',X))J(S_E(t,X))) 
\rangle_0=\cr
&\int_0^\infty dt\int_0^t dt'\langle\hat x\Lambda^{-1}(t',X)\hat x 
J(S_E(t,X))) \rangle_0+\cr
&\int_0^\infty dt\int_0^t dt'\langle\hat x\Lambda^{-1}(t',X)
\cdot\bv J(S_E(t',X))J(S_E(t,X))
\rangle_0 }\Eq(prob)$$
where $\Lambda(t,X)$ is the tangent flow generated by $S(t,X)$, \ie
$\Lambda(t,X)={\partial S(t,X)\over \partial X}$. Eq. \equ(dkaw1),
together with analogous expressions for the higher order derivatives
can be obtained easily from the formalism developed in \cita{Ru}. Note
that since the system has Lyapunov exponents $\l^u(E)>0$ and one
$\l^s(E)<0$ (at least for not too large $E$) one should expect that
$\hat x\Lambda^{-1}(t',X)\hat x\simeq e^{-\l^s(E)t}$. This makes the
convergence of the integrals in \equ(prob) very problematic. We will
return to these equations later after we consider this question from a
different point of view\annotano{ Much of the following material is
the result of discussions that the authors had with G. Gallavotti,
N. Chernov, L.-S.  Young, D. Ruelle and C. Dettman.  We are
particularly indebted to G. Gallavotti for explaining to us a possible
method to construct a Markov partition for the billiard, construction
that forms the basis of our analysis of the conjugation map.}.

It is convenient at this point to introduce a discrete time version of
our dynamical system making precise what was already sketched in the
introduction.  For every point $X$ in phase space we can define
$\tau(X)$ as the first time at which the particle collides with one of
the scatterers. Let now $\SS_E(X)=S_E(\tau(X)^+,X)$ where the ${}^+$
indicates that we choose the velocity after the collision. It is clear
that $\SS_E(X)$ restricted to the set $\TT$ of points $X=(\bq,\bv)$,
such that $\bq$ is on the surface of an obstacle and $\bv$ is directed
outward of the obstacle, defines a dynamical system (the set $\TT$ is
usually called a Poincar\'e section).  We can parametrize the
collision points $P$ in $\TT$ by two angles and an integer, \ie
$(\alpha,\beta,s)$ where $s\in\{0,1\}$ is 0 if the point is on the
obstacle with radius $0.79$ and 1 if it is on the obstacle of radius
$0.39$, $\alpha\in[-\pi,\pi]$ is the angle between the direction of
the field and the point on the scatterer at which the particle collides
and $\beta\in[-\pi/2,\pi/2]$ is the angle between the outgoing
velocity and the normal direction to the scatterer, see
Fig. \equ(Fig1).  We can also write $\TT=\TT^0\cup\TT^1$ where
$P\in\TT^s$ iff $P=(\alpha,\beta,s)$.  With a slight abuse of notation
we will still indicate our dynamical system in the new coordinates by
$\SS_E(P)$ (note that for $E=0$ the invariant measure on $\TT$ is
proportional to $\cos\beta d\beta d\alpha$).

As we already mentioned in the introduction the invariant measure
$\mu_E^+$ for $S_E(t,X)$ will induce a measure $\nu_E^+$ for $\SS(P)$
on $\TT$. The average of any function $G(X)$ with
respect to $\mu_E^+$ can be obtained from $\nu_E^+$ via:

$$\langle G(X)\rangle_E={1\over\tau^+_E}\int_\TT\GG(P)d\nu_E^+(P)\Eq(ind)$$
where $\GG(X)=\int_0^{\tau(X)}G(S_E(t,X))dt$ and
$\tau^+_E=\int_\TT\tau(P)d\nu_E^+(P)$.

A very useful tool for the discussion of regularity properties of
dynamical system in terms of external parameters consists in the
construction of a conjugating map, \ie we can look for a family of
functions $h_{E,E'}(P)$ with the property that:

$$\SS_E(h_{E,E'}(P))=h_{E,E'}(\SS_{E'}(P)).\Eq(con)$$
see \cita{BKL} for an example of such a construction for an Anosov
systems. It is easy to see that for smooth dynamical systems (\eg
Anosov or axiom A) such a family exists, at least in a small
neighborhood of any given $E$, and the average of a smooth function
has the same regularity properties as the conjugation\cita{LL}\cita{BKL}.  
In our case the existence of a smooth family $h_{E,E'}$ of conjugation
would not be enough to prove that if $\GG(P)$ is a smooth function
then $\langle\GG\rangle_E^d=\int_\TT\GG(P)d\nu_E^+(P)$ is a smooth
function of $E$. One would also have to show that the local unstable
foliation as a function of $E$ has good smoothness property
\annotano{See \cita{BKL} for aprecise discussion of this point.}. 
We will not touch this second point and focus our attention on the
conjugation, hoping that an answer to the problem of its existence
will give us enough information on the smoothness of the $\kappa(E)$.
We will try to construct such a conjugation and show that this is
impossible at least in the usual strong sense. The analysis of the
problems that one encounter in this attemped construction will lead us
to identify the possible origin of the observed behavior of
$\kappa(E)$. 

The main problem with our system is that the map $\SS_E(X)$ is only
piecewise smooth\annotano{We will use the term smooth in the vague
sense of ``sufficiently differentiable'': at the level of rigor of
the forthcoming discussion it is not worthwhile to specify exactly what
kind of regularity we need. This ambiguity should not
undermine the understandability of the following reasoning.}. In fact
$\SS_E(X)$ is smooth except at points whose image is tangent at
collision\annotano{The trajectories issuing from these points are
usually called ``grazing trajectories''},
\ie the discontinuity set $\DD^1_E$ of $\SS_E(P)$ is defined by the set of $P$
such that
$\SS_E(P)=(\alpha^{\prime},\pm \pi/2,s)$ for some $\alpha^{\prime}$. It is clear that we can divide
$\DD^1_E$ into two parts $\DD^{1,s}_E=\DD^{1}_E\cap\TT^s$
corresponding to the two obstacles.  Moreover, due to the time
reversibility of the dynamics it is easy to see that if
$(\alpha,\beta,s)\in\DD^1_E$ then $(\alpha,-\beta,s)$ is in the
discontinuity set $\DD^{-1}_E$ of $\SS_E^{-1}$. The set $\DD^{1,1}_0$
is shown in Fig. \equ(fig_d).

\dimen2=25pt \advance\dimen2 by 269pt
\dimen3=404pt
\eqfig{\dimen3}{314pt}{
\ins{358pt}{0pt}{$\hbox to20pt{\hfill ${\scriptstyle \alpha}$\hfill}$}
\ins{30pt}{294pt}{$\hbox to25pt{\hfill ${\scriptstyle \beta}$\hfill}$}
\ins{178pt}{287pt}{$\hbox to158pt{\hfill ${\scriptstyle E=0}$\hfill}$}
\ins{-4pt}{3pt}{$\hbox to 79pt{\hfill${\scriptstyle -3.14}$\hfill}$}
\ins{75pt}{3pt}{$\hbox to 79pt{\hfill${\scriptstyle -1.57}$\hfill}$}
\ins{154pt}{3pt}{$\hbox to 79pt{\hfill${\scriptstyle 0.00}$\hfill}$}
\ins{233pt}{3pt}{$\hbox to 79pt{\hfill${\scriptstyle 1.57}$\hfill}$}
\ins{313pt}{3pt}{$\hbox to 79pt{\hfill${\scriptstyle 3.14}$\hfill}$}
\ins{0pt}{25pt}{$\hbox to 32pt{\hfill${\scriptstyle -1.570}$}$}
\ins{0pt}{86pt}{$\hbox to 32pt{\hfill${\scriptstyle -0.785}$}$}
\ins{0pt}{148pt}{$\hbox to 32pt{\hfill${\scriptstyle 0.000}$}$}
\ins{0pt}{210pt}{$\hbox to 32pt{\hfill${\scriptstyle 0.785}$}$}
\ins{0pt}{272pt}{$\hbox to 32pt{\hfill${\scriptstyle 1.570}$}$}
}{disc}{Discontinuity set $\EE^{1,1}_0$ for the map $\SS_0$ on the 
obstacle of radius $0.39$: the solid lines represent points that collide 
with $\beta=\pi/2$ while the dashed lines represent points that collide 
with $\beta=-\pi/2$. The $\alpha$ has to be thought of as periodic.}{fig_d}0

This set is formed by a finite number of disjoint locally
one dimensional manifolds with possibly a finite number of bifurcations
(see Fig. \equ(fig_d)). The bifurcation points are the points $P$ where
a multiple tangency happens, \ie where the $\beta$-component of both
$\SS_E(P)$ and $\SS^2_E(P)$ is in $\{-\pi/2,\pi/2\}$. It is also easy
to see that the boundary points of these manifolds are in $\beta=\pi/2$
or $\beta=-\pi/2$ and that every branch of $\DD^{1}_E$ can be
represented by an increasing function $\alpha(\beta)$. Although
Fig. \equ(fig_d) has been generated with the help of the molecular dynamics
program described in the previous section it would not be difficult
to write analytical expressions for the manifolds in $\DD^{1}_E$.
We note that from eq.\equ(ind) we get that $\langle J(X)\rangle_E=
\langle \JJ(P)\rangle^d_E/\tau^+_E$ where $\JJ(P)$ is not a smooth
function on $\TT$. This should not be a problem considering that
$\JJ$, as any function $\GG(P)$ obtained from a smooth function $G(X)$
on phase space in the same way, is discontinuous exaclty on $\DD^1_E$
and smooth everywhere else.

It is also easy to observe that $\II^1_E=\DD^{-1}_E\cap\DD^1_E$ is a
discrete set with a finite number $I_E$ of points $P_i$. We can assume
that $I_E$ is a piecewise constant function of $E$ and one can write
$P_i=P_i(E)$ for every $P_i\in\II^1_E$ with $P_i(E)$ defined and
smooth on every open interval on which $I_E$ is constant.  Moreover
the points $P_i(E)$, together with the bifurcation points, divide the
manifolds to which they belong into connected segments and these
segments form the sides of polygons (generically squares and triangles
at this level) that form a partition of $\TT$. Again we can
assume that, for $E$ in a small enough interval, these segments and
polygons can be smoothly parametrized. Fig. \equ(fig_dd) presents a
snapshot of this situation for $E=0$.
%
\dimen2=25pt \advance\dimen2 by 269pt
\dimen3=404pt
\eqfig{\dimen3}{314pt}{
\ins{358pt}{0pt}{$\hbox to20pt{\hfill ${\scriptstyle \alpha}$\hfill}$}
\ins{30pt}{294pt}{$\hbox to25pt{\hfill ${\scriptstyle \beta}$\hfill}$}
\ins{178pt}{287pt}{$\hbox to158pt{\hfill ${\scriptstyle E=0}$\hfill}$}
\ins{-4pt}{3pt}{$\hbox to 79pt{\hfill${\scriptstyle -3.14}$\hfill}$}
\ins{75pt}{3pt}{$\hbox to 79pt{\hfill${\scriptstyle -1.57}$\hfill}$}
\ins{154pt}{3pt}{$\hbox to 79pt{\hfill${\scriptstyle 0.00}$\hfill}$}
\ins{233pt}{3pt}{$\hbox to 79pt{\hfill${\scriptstyle 1.57}$\hfill}$}
\ins{313pt}{3pt}{$\hbox to 79pt{\hfill${\scriptstyle 3.14}$\hfill}$}
\ins{0pt}{25pt}{$\hbox to 32pt{\hfill${\scriptstyle -1.570}$}$}
\ins{0pt}{86pt}{$\hbox to 32pt{\hfill${\scriptstyle -0.785}$}$}
\ins{0pt}{148pt}{$\hbox to 32pt{\hfill${\scriptstyle 0.000}$}$}
\ins{0pt}{210pt}{$\hbox to 32pt{\hfill${\scriptstyle 0.785}$}$}
\ins{0pt}{272pt}{$\hbox to 32pt{\hfill${\scriptstyle 1.570}$}$}
}{discd}{Discontinuity set $\DD^{1,1}_0$ and $\DD^{-1,1}_0$ for the 
map $\SS_E(P)$ on the obstacle of radius $0.39$}{fig_dd}0
It is clear that any conjugating map $h_{E,E'}$ has to map
$\DD^1_E$ to $\DD^{1}_{E'}$ and $\DD^{-1}_{E'}$ to $\DD^{-1}_{E'}$. A
possible first approximation of such an $h_{E,E'}$
consists in defining $h^{(1)}_{E,E'}(P_i(E))\big|_{\II_E^1}=P_i(E')$. It
is clear that such a map can be extended to a smooth map
$h^{(1)}_{E,E'}:\TT\to\TT$ such that $h^{(1)}_{E,E'}(P)=P+(E'-E)\delta
h^{(1)}_{E,E'}(P)$ where $\delta
h^{(1)}_{E,E'}$ is some smooth function and $h^{(1)}_{E,E'}(\DD^{\s}_{E})=\DD^{\s}_{E'}$ for
$\s=\pm 1$.

Although the map $h^{(1)}_{E,E'}$ is surely not a conjugation we can
consider it as a starting point and reproduce the above construction
using the sets of discontinuities of $\SS^2_E(P)$ and of
$\SS^{-2}_E(P)$, $\DD^2_{E}$ and $\DD^{-2}_{E}$ respectively,
obtaining a new map \ $h^{(2)}_{E,E'}$. Observe that
$\DD^2_{E}=\DD^1_{E}\cup \tilde \DD^2_{E}$ where $P\in\tilde\DD^2_{E}$
iff the $\beta$-component of $\SS^2_E(P)$ is in $\{-\pi/2,\pi/2\}$.
An example of the set $\tilde \DD^2_{E}\cup\tilde \DD^{-2}_{E}$ is
given in Fig. \equ(fig_d2)

\dimen2=25pt \advance\dimen2 by 269pt
\dimen3=404pt
\eqfig{\dimen3}{314pt}{
\ins{358pt}{0pt}{$\hbox to20pt{\hfill ${\scriptstyle \alpha}$\hfill}$}
\ins{30pt}{294pt}{$\hbox to25pt{\hfill ${\scriptstyle \beta}$\hfill}$}
\ins{178pt}{287pt}{$\hbox to158pt{\hfill ${\scriptstyle E=0}$\hfill}$}
\ins{-4pt}{3pt}{$\hbox to 79pt{\hfill${\scriptstyle -3.14}$\hfill}$}
\ins{75pt}{3pt}{$\hbox to 79pt{\hfill${\scriptstyle -1.57}$\hfill}$}
\ins{154pt}{3pt}{$\hbox to 79pt{\hfill${\scriptstyle 0.00}$\hfill}$}
\ins{233pt}{3pt}{$\hbox to 79pt{\hfill${\scriptstyle 1.57}$\hfill}$}
\ins{313pt}{3pt}{$\hbox to 79pt{\hfill${\scriptstyle 3.14}$\hfill}$}
\ins{0pt}{23pt}{$\hbox to 32pt{\hfill${\scriptstyle -1.570}$}$}
\ins{0pt}{86pt}{$\hbox to 32pt{\hfill${\scriptstyle -0.785}$}$}
\ins{0pt}{148pt}{$\hbox to 32pt{\hfill${\scriptstyle 0.000}$}$}
\ins{0pt}{211pt}{$\hbox to 32pt{\hfill${\scriptstyle 0.785}$}$}
\ins{0pt}{273pt}{$\hbox to 32pt{\hfill${\scriptstyle 1.570}$}$}
}{disc2}{Discontinuity set $\DD^{2,1}_0$ and
$\DD^{-2,1}_0$ for the map $\SS_E(P)$ on the obstacle of radius
$0.395$}{fig_d2}0

It is well known that the sets $\DD^{N}_E$ will become dense as
$N\to\infty$ so one can hope that the above construction will
create a sequence of functions $h^{(N)}_{E,E'}$ eventually converging
to a real conjugation when $N\to\infty$. Moreover, considering that
all the so constructed approximations are smooth one can hope that also
the limit will be smooth. There is nevertheless a major problem that makes
such a convergence impossible, at least in this first naive meaning,
as we will see in what follows.

Let us again consider our first approximation $h^{(1)}_{E,E'}$. It is
easy to realize that we can associate to each of the connected regions
in which $\DD^{1,s}_{E}$ divides $\TT^s$ a pair of integer
$\bc=(n_x,n_y)$ indicating the difference in coordinates between the
center of the obstacle with which the particle collides and the one
from which it starts. From now on we will see the dynamics as taking
place on the universal covering of the torus and we will assume that
the origin coincides with the center of one of the scatterers of radius
$0.79$. We will denote the set of this pair of integers by $\tilde
D^{1,s}_E$ and call it the set of possible ``forward histories of lenght
one'' for reasons that will become clear in what follows.  Observe that
the map $\SS_E$ is continuous from one side at every point on
$\DD^{1}_{E}$. This allows us to define the set $D^{1,s}_E\subset
\tilde D^{1,s}_E$ of the pair of integers associated with points in
$\DD^{1,s}_{E}$ (these represent the obstacles that can be grazed
starting from a given obstacle). In Fig. \equ(figc) we show, in the two
upper pictures, the set $D^{1,1}_E$ for $E=0$ and $E=0.45$. One can
realize that two new obstacles can be grazed at a field $E=0.45$ whereas they
were not at a field $E=0$. The two lower pictures show the
corresponding appearence of a new branch of $\DD^{1}_{E}$ and the fact
that a new connected component of $\TT^1\backslash\DD^{1}_{E}$ is
created. The number on the figures are meant to help associate
corresponding regions. The real modification of the set $D^{1,1}_E$
takes place at some value of $E\in(0.375,0.4)$ but we show the
situation at $E=0.45$ because it would otherwise be difficult to
observe it due to the small size of the region denoted by $5$ in the
lower right picture in Fig. \equ(figc). It is interesting to note that
this change takes place in one of the intervals where we located a
possible discontinuity of the derivative of $\kappa(E)$.  We believe
that this phenomenon is at the origin of those discontinuities and we
will argue in this sense in what follows.

\dimen2=25pt \advance\dimen2 by 126pt
\dimen3=219pt
\dimenfor{\dimen3}{166pt}
\eqfigfor{
\ins{199pt}{0pt}{$\hbox to20pt{\hfill ${\scriptstyle n_x}$\hfill}$}
\ins{24pt}{\dimen2}{$\hbox to25pt{\hfill ${\scriptstyle n_y}$\hfill}$}
\ins{107pt}{144pt}{$\hbox to87pt{\hfill ${\scriptstyle E=0.0}$\hfill}$}
\ins{18pt}{3pt}{$\hbox to 39pt{\hfill${\scriptstyle -2}$\hfill}$}
\ins{57pt}{3pt}{$\hbox to 39pt{\hfill${\scriptstyle -1}$\hfill}$}
\ins{97pt}{3pt}{$\hbox to 39pt{\hfill${\scriptstyle 0}$\hfill}$}
\ins{136pt}{3pt}{$\hbox to 39pt{\hfill${\scriptstyle 1}$\hfill}$}
\ins{176pt}{3pt}{$\hbox to 39pt{\hfill${\scriptstyle 2}$\hfill}$}
\ins{0pt}{19pt}{$\hbox to 26pt{\hfill${\scriptstyle -3}$}$}
\ins{0pt}{38pt}{$\hbox to 26pt{\hfill${\scriptstyle -2}$}$}
\ins{0pt}{58pt}{$\hbox to 26pt{\hfill${\scriptstyle -1}$}$}
\ins{0pt}{77pt}{$\hbox to 26pt{\hfill${\scriptstyle 0}$}$}
\ins{0pt}{96pt}{$\hbox to 26pt{\hfill${\scriptstyle 1}$}$}
\ins{0pt}{115pt}{$\hbox to 26pt{\hfill${\scriptstyle 2}$}$}
\ins{0pt}{134pt}{$\hbox to 26pt{\hfill${\scriptstyle 3}$}$}
}{
\ins{199pt}{0pt}{$\hbox to20pt{\hfill ${\scriptstyle n_x}$\hfill}$}
\ins{24pt}{\dimen2}{$\hbox to25pt{\hfill ${\scriptstyle n_y}$\hfill}$}
\ins{107pt}{144pt}{$\hbox to87pt{\hfill ${\scriptstyle E=0.45}$\hfill}$}
\ins{18pt}{3pt}{$\hbox to 39pt{\hfill${\scriptstyle -2}$\hfill}$}
\ins{57pt}{3pt}{$\hbox to 39pt{\hfill${\scriptstyle -1}$\hfill}$}
\ins{97pt}{3pt}{$\hbox to 39pt{\hfill${\scriptstyle 0}$\hfill}$}
\ins{136pt}{3pt}{$\hbox to 39pt{\hfill${\scriptstyle 1}$\hfill}$}
\ins{176pt}{3pt}{$\hbox to 39pt{\hfill${\scriptstyle 2}$\hfill}$}
\ins{0pt}{19pt}{$\hbox to 26pt{\hfill${\scriptstyle -3}$}$}
\ins{0pt}{38pt}{$\hbox to 26pt{\hfill${\scriptstyle -2}$}$}
\ins{0pt}{58pt}{$\hbox to 26pt{\hfill${\scriptstyle -1}$}$}
\ins{0pt}{77pt}{$\hbox to 26pt{\hfill${\scriptstyle 0}$}$}
\ins{0pt}{96pt}{$\hbox to 26pt{\hfill${\scriptstyle 1}$}$}
\ins{0pt}{115pt}{$\hbox to 26pt{\hfill${\scriptstyle 2}$}$}
\ins{0pt}{134pt}{$\hbox to 26pt{\hfill${\scriptstyle 3}$}$}
}{
\ins{199pt}{0pt}{$\hbox to20pt{\hfill ${\scriptstyle \alpha}$\hfill}$}
\ins{24pt}{\dimen2}{$\hbox to25pt{\hfill ${\scriptstyle \beta}$\hfill}$}
\ins{107pt}{144pt}{$\hbox to87pt{\hfill ${\scriptstyle E=0.0}$\hfill}$}
\ins{18pt}{3pt}{$\hbox to 39pt{\hfill${\scriptstyle 2.2}$\hfill}$}
\ins{57pt}{3pt}{$\hbox to 39pt{\hfill${\scriptstyle 2.4}$\hfill}$}
\ins{97pt}{3pt}{$\hbox to 39pt{\hfill${\scriptstyle 2.6}$\hfill}$}
\ins{136pt}{3pt}{$\hbox to 39pt{\hfill${\scriptstyle 2.8}$\hfill}$}
\ins{176pt}{3pt}{$\hbox to 39pt{\hfill${\scriptstyle 3.0}$\hfill}$}
\ins{0pt}{19pt}{$\hbox to 26pt{\hfill${\scriptstyle -1.50}$}$}
\ins{0pt}{45pt}{$\hbox to 26pt{\hfill${\scriptstyle -1.30}$}$}
\ins{0pt}{70pt}{$\hbox to 26pt{\hfill${\scriptstyle -1.10}$}$}
\ins{0pt}{96pt}{$\hbox to 26pt{\hfill${\scriptstyle -0.90}$}$}
\ins{0pt}{122pt}{$\hbox to 26pt{\hfill${\scriptstyle -0.70}$}$}
\ins{38pt}{32pt}{$\hbox to 26pt{\hfill${\scriptstyle 1}$}$}
\ins{76pt}{32pt}{$\hbox to 26pt{\hfill${\scriptstyle 2}$}$}
\ins{88pt}{40pt}{$\hbox to 26pt{\hfill${\scriptstyle 3}$}$}
\ins{100pt}{70pt}{$\hbox to 26pt{\hfill${\scriptstyle 4}$}$}
}{
\ins{199pt}{0pt}{$\hbox to20pt{\hfill ${\scriptstyle \alpha}$\hfill}$}
\ins{24pt}{\dimen2}{$\hbox to25pt{\hfill ${\scriptstyle \beta}$\hfill}$}
\ins{107pt}{144pt}{$\hbox to87pt{\hfill ${\scriptstyle E=0.45}$\hfill}$}
\ins{18pt}{3pt}{$\hbox to 39pt{\hfill${\scriptstyle 2.2}$\hfill}$}
\ins{57pt}{3pt}{$\hbox to 39pt{\hfill${\scriptstyle 2.4}$\hfill}$}
\ins{97pt}{3pt}{$\hbox to 39pt{\hfill${\scriptstyle 2.6}$\hfill}$}
\ins{136pt}{3pt}{$\hbox to 39pt{\hfill${\scriptstyle 2.8}$\hfill}$}
\ins{176pt}{3pt}{$\hbox to 39pt{\hfill${\scriptstyle 3.0}$\hfill}$}
\ins{0pt}{19pt}{$\hbox to 26pt{\hfill${\scriptstyle -1.50}$}$}
\ins{0pt}{45pt}{$\hbox to 26pt{\hfill${\scriptstyle -1.30}$}$}
\ins{0pt}{70pt}{$\hbox to 26pt{\hfill${\scriptstyle -1.10}$}$}
\ins{0pt}{96pt}{$\hbox to 26pt{\hfill${\scriptstyle -0.90}$}$}
\ins{0pt}{122pt}{$\hbox to 26pt{\hfill${\scriptstyle -0.70}$}$}
\ins{38pt}{32pt}{$\hbox to 26pt{\hfill${\scriptstyle 1}$}$}
\ins{60pt}{85pt}{$\hbox to 26pt{\hfill${\scriptstyle 2}$}$}
\ins{13pt}{130pt}{$\hbox to 26pt{\hfill${\scriptstyle 3}$}$}
\ins{120pt}{90pt}{$\hbox to 26pt{\hfill${\scriptstyle 4}$}$}
\ins{166pt}{17pt}{$\hbox to 26pt{\hfill${\scriptstyle 5}$}$}
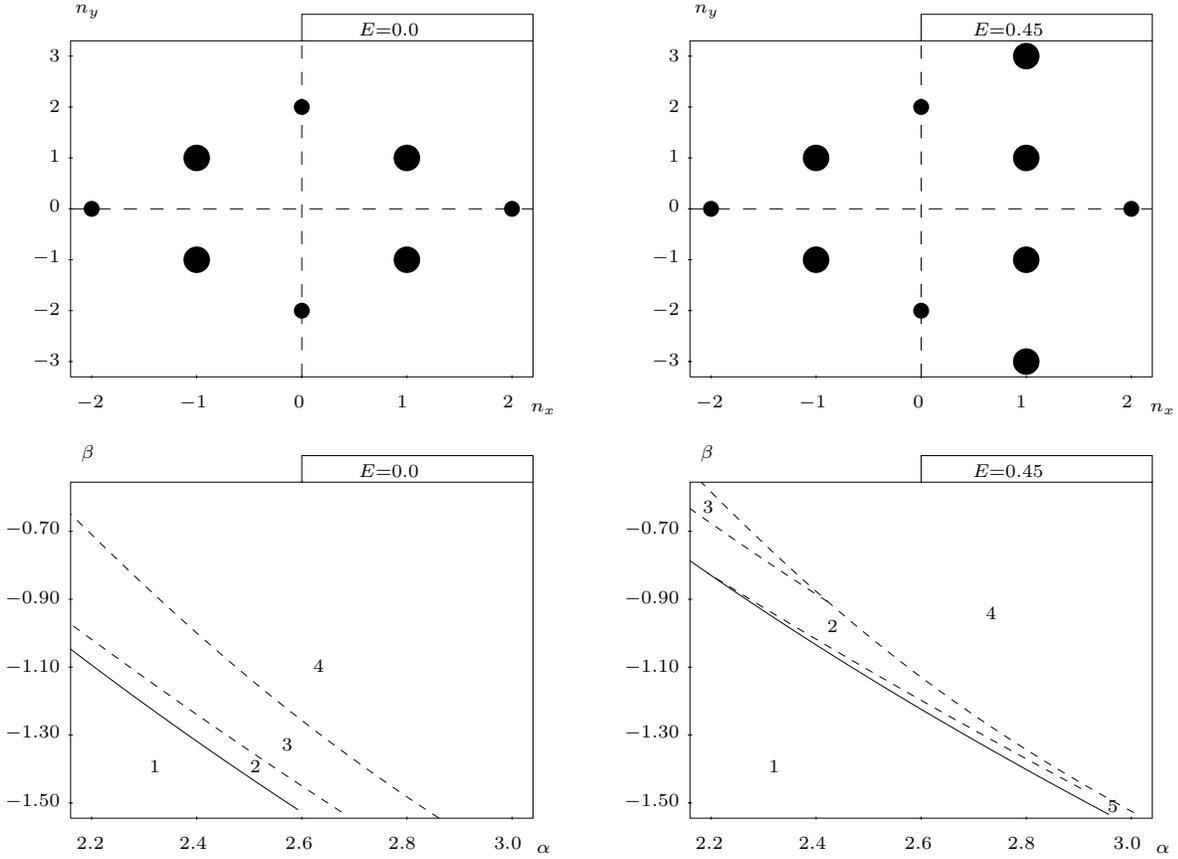
}{centri_1}{$D^{1,1}_E$ for $E=0$ and $E=0.45$ with a detailed wiev of a 
part of the 
corresponding set $\EE_E^{1,1}$ showing the appearence of a new branch and a 
new connected component in $\TT^1\backslash\EE_E{1,1}$ (numbers represent 
corresponding regions). Although the sizes are not 
to scale a larger point stands for an obstacle of radius $R_2$.
}{figc}

Let us call $E_d$ the exact value of $E$ at which the new possible
histories appear. We can still construct our map $h^{(1)}_{E_d-\d
E,E_d+\d E}$, for a very small $\d E$, mapping the regions numbered
from 1 to 4 in Fig. \equ(fig_d) for $E=E_d-\d E$ in to the
corresponding ones for $E=E_d+\d E$. This would create the first
approximation $h^{(1)}_{E_d-\d E,E_d+\d E}$ to what we can call a
``partial conjugation'' $h_{E_d-\d E,E_d+\d E}$ to be obtained as a
limit of functions $h^{(N)}_{E_d-\d E,E_d+\d E}$ constructed iterating
the above scheme.  Using this function to compute the average of a
smooth function on $\TT$ as a function of $E$ would give a smooth
result but we will commit an error due to the missed region
corresponding to the new possible history (region 5 in
Fig. \equ(fig_d)). It is easy to realize that this region will have a
size proportional to $\sqrt{\d E}$ and thus an area proportional to $\d
E$. We can assume, as a first approximation, that the error committed
in neglecting such a region is proportional to this area which would
give a discontinuity in the first derivative.

We will now check if we can use this idea to account for the other
points of discontinuity we have indicated in the previous
section. First of all we observe that something very similar to what
we discussed above happens in the interval $(0.325,0.35)$ with the
role of $D^{1,1}_E$ played by $D^{1,0}_E$, as shown in
Fig. \equ(fig_c2).

\dimen2=25pt \advance\dimen2 by 126pt
\dimen3=212pt
\eqfigbis{\dimen3}{166pt}{
\ins{190pt}{0pt}{$\hbox to20pt{\hfill ${\scriptstyle n_x}$\hfill}$}
\ins{12pt}{151pt}{$\hbox to25pt{\hfill ${\scriptstyle n_y}$\hfill}$}
\ins{107pt}{144pt}{$\hbox to87pt{\hfill ${\scriptstyle E=0.0}$\hfill}$}
\ins{16pt}{3pt}{$\hbox to 19pt{\hfill${\scriptstyle -4}$\hfill}$}
\ins{35pt}{3pt}{$\hbox to 19pt{\hfill${\scriptstyle -3}$\hfill}$}
\ins{55pt}{3pt}{$\hbox to 19pt{\hfill${\scriptstyle -2}$\hfill}$}
\ins{75pt}{3pt}{$\hbox to 19pt{\hfill${\scriptstyle -1}$\hfill}$}
\ins{95pt}{3pt}{$\hbox to 19pt{\hfill${\scriptstyle 0}$\hfill}$}
\ins{115pt}{3pt}{$\hbox to 19pt{\hfill${\scriptstyle 1}$\hfill}$}
\ins{134pt}{3pt}{$\hbox to 19pt{\hfill${\scriptstyle 2}$\hfill}$}
\ins{154pt}{3pt}{$\hbox to 19pt{\hfill${\scriptstyle 3}$\hfill}$}
\ins{174pt}{3pt}{$\hbox to 19pt{\hfill${\scriptstyle 4}$\hfill}$}
\ins{0pt}{19pt}{$\hbox to 14pt{\hfill${\scriptstyle -4}$}$}
\ins{0pt}{34pt}{$\hbox to 14pt{\hfill${\scriptstyle -3}$}$}
\ins{0pt}{48pt}{$\hbox to 14pt{\hfill${\scriptstyle -2}$}$}
\ins{0pt}{62pt}{$\hbox to 14pt{\hfill${\scriptstyle -1}$}$}
\ins{0pt}{77pt}{$\hbox to 14pt{\hfill${\scriptstyle 0}$}$}
\ins{0pt}{91pt}{$\hbox to 14pt{\hfill${\scriptstyle 1}$}$}
\ins{0pt}{106pt}{$\hbox to 14pt{\hfill${\scriptstyle 2}$}$}
\ins{0pt}{120pt}{$\hbox to 14pt{\hfill${\scriptstyle 3}$}$}
\ins{0pt}{134pt}{$\hbox to 14pt{\hfill${\scriptstyle 4}$}$}
}{\ins{190pt}{0pt}{$\hbox to20pt{\hfill ${\scriptstyle n_x}$\hfill}$}
\ins{12pt}{151pt}{$\hbox to25pt{\hfill ${\scriptstyle n_y}$\hfill}$}
\ins{107pt}{144pt}{$\hbox to87pt{\hfill ${\scriptstyle E=0.35}$\hfill}$}
\ins{16pt}{3pt}{$\hbox to 19pt{\hfill${\scriptstyle -4}$\hfill}$}
\ins{35pt}{3pt}{$\hbox to 19pt{\hfill${\scriptstyle -3}$\hfill}$}
\ins{55pt}{3pt}{$\hbox to 19pt{\hfill${\scriptstyle -2}$\hfill}$}
\ins{75pt}{3pt}{$\hbox to 19pt{\hfill${\scriptstyle -1}$\hfill}$}
\ins{95pt}{3pt}{$\hbox to 19pt{\hfill${\scriptstyle 0}$\hfill}$}
\ins{115pt}{3pt}{$\hbox to 19pt{\hfill${\scriptstyle 1}$\hfill}$}
\ins{134pt}{3pt}{$\hbox to 19pt{\hfill${\scriptstyle 2}$\hfill}$}
\ins{154pt}{3pt}{$\hbox to 19pt{\hfill${\scriptstyle 3}$\hfill}$}
\ins{174pt}{3pt}{$\hbox to 19pt{\hfill${\scriptstyle 4}$\hfill}$}
\ins{0pt}{19pt}{$\hbox to 14pt{\hfill${\scriptstyle -4}$}$}
\ins{0pt}{34pt}{$\hbox to 14pt{\hfill${\scriptstyle -3}$}$}
\ins{0pt}{48pt}{$\hbox to 14pt{\hfill${\scriptstyle -2}$}$}
\ins{0pt}{62pt}{$\hbox to 14pt{\hfill${\scriptstyle -1}$}$}
\ins{0pt}{77pt}{$\hbox to 14pt{\hfill${\scriptstyle 0}$}$}
\ins{0pt}{91pt}{$\hbox to 14pt{\hfill${\scriptstyle 1}$}$}
\ins{0pt}{106pt}{$\hbox to 14pt{\hfill${\scriptstyle 2}$}$}
\ins{0pt}{120pt}{$\hbox to 14pt{\hfill${\scriptstyle 3}$}$}
\ins{0pt}{134pt}{$\hbox to 14pt{\hfill${\scriptstyle 4}$}$}
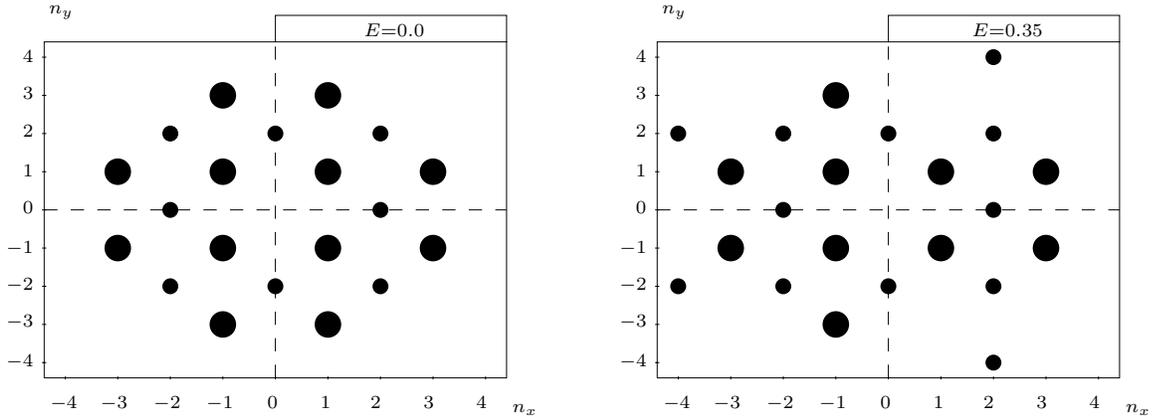
}{centri_2}{$D^{0,1}_E$ for $E=0$ and $E=0.35$.}{fig_c2}

Although it seems to agree quite well with the numerical observations
this picture is still very partial. In fact it can happen that the set
$\II^{1}_E=\DD^{-1}_E\cap\DD^1_E$ changes structure although the sets
$\DD^{\pm 1}_E$ remain structurally identical. It is possible to see
from Fig. \equ(fig_dd) that almost all the points in $\II^1_E$ are near
bifurcation points for $\DD^{\pm 1}_E$. Looking carefully at the
picture (see Fig. \equ(fig_c2)) one realizes that there are indeed 4
intesection points that delimit a small square. We observe that, as we
did before, we can associate with every polygon in $\TT$ bounded by
$\DD^{-1,s}_E\cup\DD^{1,s}_E$ a sequence of length 2
$(\bc_{-1},\bc_1)$ of pairs of integers $\bc_i=(n_{x,i},n_{y,i})$
representing the center on which the particle collides forward and
backward in time. We can call the set of all these pairs the possible
``symmetric history of length two'' and denote it by $M^{1}_E$.  If
one follows the evolution of the above mentioned square as a function
of the field one realizes that at a field in $(0.05,0.075)$ it
disappears (see Fig. \equ(fig_c2)).  We can again reason as before to
deduce that this phenomenon will give rise to a discontinuity of the
first derivative. The possible definition of the map $h^{(1)}_{E,E'}$
in this situation appears more complex and we will not discuss it here
leaving this question and a formalization of the above picture to a
forthcoming study.

\dimen2=25pt \advance\dimen2 by 126pt
\dimen3=214pt
\eqfigbis{\dimen3}{166pt}{
\ins{190pt}{0pt}{$\hbox to20pt{\hfill ${\scriptstyle \alpha}$\hfill}$}
\ins{30pt}{142pt}{$\hbox to25pt{\hfill ${\scriptstyle \beta}$\hfill}$}
\ins{93pt}{135pt}{$\hbox to73pt{\hfill ${\scriptstyle E=0.0}$\hfill}$}
\ins{25pt}{3pt}{$\hbox to 24pt{\hfill${\scriptstyle -0.03}$\hfill}$}
\ins{49pt}{3pt}{$\hbox to 24pt{\hfill${\scriptstyle -0.02}$\hfill}$}
\ins{73pt}{3pt}{$\hbox to 24pt{\hfill${\scriptstyle -0.01}$\hfill}$}
\ins{97pt}{3pt}{$\hbox to 24pt{\hfill${\scriptstyle 0.00}$\hfill}$}
\ins{121pt}{3pt}{$\hbox to 24pt{\hfill${\scriptstyle 0.01}$\hfill}$}
\ins{145pt}{3pt}{$\hbox to 24pt{\hfill${\scriptstyle 0.02}$\hfill}$}
\ins{169pt}{3pt}{$\hbox to 24pt{\hfill${\scriptstyle 0.03}$\hfill}$}
\ins{0pt}{15pt}{$\hbox to 32pt{\hfill${\scriptstyle -0.28}$}$}
\ins{0pt}{53pt}{$\hbox to 32pt{\hfill${\scriptstyle -0.27}$}$}
\ins{0pt}{91pt}{$\hbox to 32pt{\hfill${\scriptstyle -0.26}$}$}
\ins{0pt}{130pt}{$\hbox to 32pt{\hfill${\scriptstyle -0.25}$}$}
}{\ins{190pt}{0pt}{$\hbox to20pt{\hfill ${\scriptstyle \alpha}$\hfill}$}
\ins{30pt}{142pt}{$\hbox to25pt{\hfill ${\scriptstyle \beta}$\hfill}$}
\ins{93pt}{135pt}{$\hbox to73pt{\hfill ${\scriptstyle E=0.075}$\hfill}$}
\ins{25pt}{3pt}{$\hbox to 24pt{\hfill${\scriptstyle -0.03}$\hfill}$}
\ins{49pt}{3pt}{$\hbox to 24pt{\hfill${\scriptstyle -0.02}$\hfill}$}
\ins{73pt}{3pt}{$\hbox to 24pt{\hfill${\scriptstyle -0.01}$\hfill}$}
\ins{97pt}{3pt}{$\hbox to 24pt{\hfill${\scriptstyle 0.00}$\hfill}$}
\ins{121pt}{3pt}{$\hbox to 24pt{\hfill${\scriptstyle 0.01}$\hfill}$}
\ins{145pt}{3pt}{$\hbox to 24pt{\hfill${\scriptstyle 0.02}$\hfill}$}
\ins{169pt}{3pt}{$\hbox to 24pt{\hfill${\scriptstyle 0.03}$\hfill}$}
\ins{0pt}{15pt}{$\hbox to 32pt{\hfill${\scriptstyle -0.31}$}$}
\ins{0pt}{34pt}{$\hbox to 32pt{\hfill${\scriptstyle -0.30}$}$}
\ins{0pt}{53pt}{$\hbox to 32pt{\hfill${\scriptstyle -0.29}$}$}
\ins{0pt}{72pt}{$\hbox to 32pt{\hfill${\scriptstyle -0.28}$}$}
\ins{0pt}{91pt}{$\hbox to 32pt{\hfill${\scriptstyle -0.27}$}$}
\ins{0pt}{111pt}{$\hbox to 32pt{\hfill${\scriptstyle -0.26}$}$}
\ins{0pt}{130pt}{$\hbox to 32pt{\hfill${\scriptstyle -0.25}$}$}
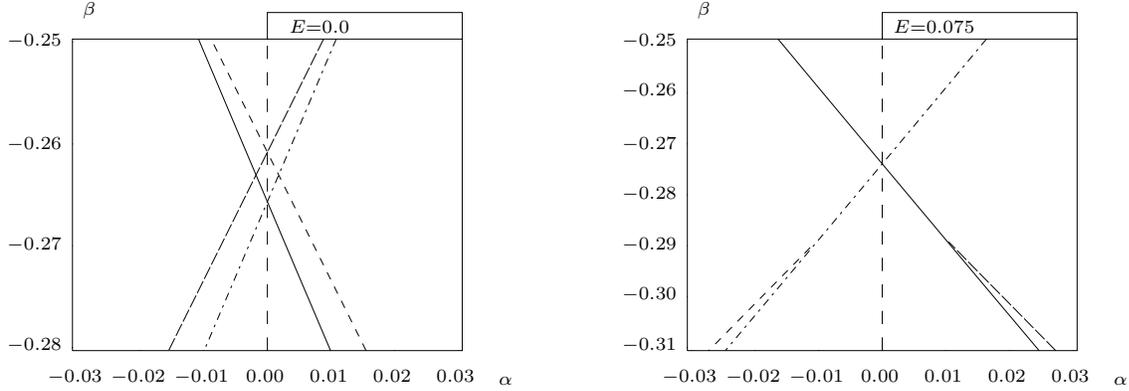
}{mar}{Detail of $\DD^{1,1}_E\cup\DD^{-1,1}_E$ showing the change of
$I^1_E$ although $\DD^{s,1}_E$ changes smoothly
}{fig_m}

We observe here that the above considerations can be iterated. We can
consider the set $D^{N,s}_E$ formed by sequences of length $N$,
$(\bc_1,\dots,\bc_N)$ of pairs of integers representing the possible
consecutive collisions that a trajectory experiences starting from the
obstacle $s$ (we can call it the set of possible ``forward histories
of length N''). With a reasoning analogous to the one above we can
expect a discontinuity of the first derivative when one of this set
changes.  It is easy to see that, as $N$ grows, the number of
histories in the set $D^{N}_E$ increases exponentially and, due to the
instability of the dynamics, new possible histories will appear or
disappear very often.  On the other hand each of the regions in which
$\TT$ is divided by the set $\DD^{N}_E$ is exponentially small in $N$
so we can expect them to create exponentially small discontinuities of
the first derivative. A similar argument holds for the set $\II^{N}_E$
and the set of symmetric histories of length $2N$. This suggests that
one should be able to define a ``partial conjugation between $E$ and
$E'$'' mapping $\TT\backslash\Sigma_{E,E'}$ to
$\TT\backslash\Sigma_{E',E}$ where $\Sigma_{E',E}$ and $\Sigma_{E,E'}$
are two sets with area proportional to $|E'-E|$ and a very complex
structure, possibly fractal.  Considering our comments after
eq.\equ(con) this would suggests that $\kappa(E)$, as well as the
average of any other smooth function of $X$, is not a smooth function
of $E$, \eg it has a dense set of discontinuities. If one accept the
above picture on the structure of these discontinuities one can
still conjecture that the function $\kappa(E)$ is Lipschitz
continuous in $E$, \ie $|\kappa(E)-\kappa(E')|\le C |E-E'|$, 
for a suitable constant $C$. This would imply that $\kappa(E)$ is 
differentiable almost everywhere with respect to the Lebesgue measure. 

We note here that in the interval $(0.25,0.275)$
several of the above phenomena take place (\ie $D^{2}_E$ and $D^{3}_E$
change and a symmetric history of length two disappears) so that a
precise discussion of what happens is not possible at this point.

The above picture may seem unnatural and without hope of rigorous
formalization and practical use.  In this respect it is interesting to
note that one can consider the polygons that are rectangles in Fig.
\equ(fig_dd) as part of a Markov (we use this notion
in a much weaker sense than the usual meaning of a partition of $\TT$,
in a possibly denumerable set of rectangles whose boundaries are
mapped covariantly one to the other for a finite number of
iterations). Indeed not all of them will have the necessary property
of covariance but it is possible to give a theoretically simple
algorithm to detect which can be considered as part of a Markov
partition. The part of phase space not covered by these squares should
be analyzed using the discontinuity set of $\SS^2_E$. This will permit
to obtain more squares and to cover a bigger portion of phase
space. Clearly this algorithm can be iterated and, hopefully, will
give rise to a complete Markov partition. In this situation our
partial conjugation map is obtained by simply matching points with the
same symbolic history. Although the above description seems quite
vague we think that the algorithm can be explicitly written and
implemented on a computer. We hope to come back to this point in a
forthcoming work.

We conclude this section with two remarks:

\item{$\bullet$}One is typically used to think of conjugation as
mapping the unstable manifold of one system into the unstable manifold
of the other and similarly for the stable. Our proposed map, in some
sense really does this. In fact the set $\DD^N_E$ will get more and more
parallel to the unstable local foliation of our system. An analogous
comment holds for the Markov partition.

\dimen2=25pt \advance\dimen2 by 269pt
\dimen3=392pt
\eqfig{\dimen3}{314pt}{
\ins{372pt}{0pt}{$\hbox to20pt{\hfill ${\scriptstyle t}$\hfill}$}
\ins{18pt}{\dimen2}{$\hbox to25pt{\hfill ${\scriptstyle d_E(t)}$\hfill}$}
\ins{194pt}{287pt}{$\hbox to174pt{\hfill ${\scriptstyle E=0.0,\;0.025,\;0.05,\;0.075}$\hfill}$}
\ins{24pt}{3pt}{$\hbox to 31pt{\hfill${\scriptstyle 0.0}$\hfill}$}
\ins{55pt}{3pt}{$\hbox to 31pt{\hfill${\scriptstyle 0.5}$\hfill}$}
\ins{87pt}{3pt}{$\hbox to 31pt{\hfill${\scriptstyle 1.0}$\hfill}$}
\ins{119pt}{3pt}{$\hbox to 31pt{\hfill${\scriptstyle 1.5}$\hfill}$}
\ins{150pt}{3pt}{$\hbox to 31pt{\hfill${\scriptstyle 2.0}$\hfill}$}
\ins{182pt}{3pt}{$\hbox to 31pt{\hfill${\scriptstyle 2.5}$\hfill}$}
\ins{214pt}{3pt}{$\hbox to 31pt{\hfill${\scriptstyle 3.0}$\hfill}$}
\ins{245pt}{3pt}{$\hbox to 31pt{\hfill${\scriptstyle 3.5}$\hfill}$}
\ins{277pt}{3pt}{$\hbox to 31pt{\hfill${\scriptstyle 4.0}$\hfill}$}
\ins{309pt}{3pt}{$\hbox to 31pt{\hfill${\scriptstyle 4.5}$\hfill}$}
\ins{340pt}{3pt}{$\hbox to 31pt{\hfill${\scriptstyle 5.0}$\hfill}$}
\ins{0pt}{26pt}{$\hbox to 20pt{\hfill${\scriptstyle -11}$}$}
\ins{0pt}{49pt}{$\hbox to 20pt{\hfill${\scriptstyle -10}$}$}
\ins{0pt}{72pt}{$\hbox to 20pt{\hfill${\scriptstyle -9}$}$}
\ins{0pt}{96pt}{$\hbox to 20pt{\hfill${\scriptstyle -8}$}$}
\ins{0pt}{119pt}{$\hbox to 20pt{\hfill${\scriptstyle -7}$}$}
\ins{0pt}{142pt}{$\hbox to 20pt{\hfill${\scriptstyle -6}$}$}
\ins{0pt}{166pt}{$\hbox to 20pt{\hfill${\scriptstyle -5}$}$}
\ins{0pt}{189pt}{$\hbox to 20pt{\hfill${\scriptstyle -4}$}$}
\ins{0pt}{212pt}{$\hbox to 20pt{\hfill${\scriptstyle -3}$}$}
\ins{0pt}{236pt}{$\hbox to 20pt{\hfill${\scriptstyle -2}$}$}
\ins{0pt}{259pt}{$\hbox to 20pt{\hfill${\scriptstyle -1}$}$}
\ins{0pt}{282pt}{$\hbox to 20pt{\hfill${\scriptstyle 0}$}$}
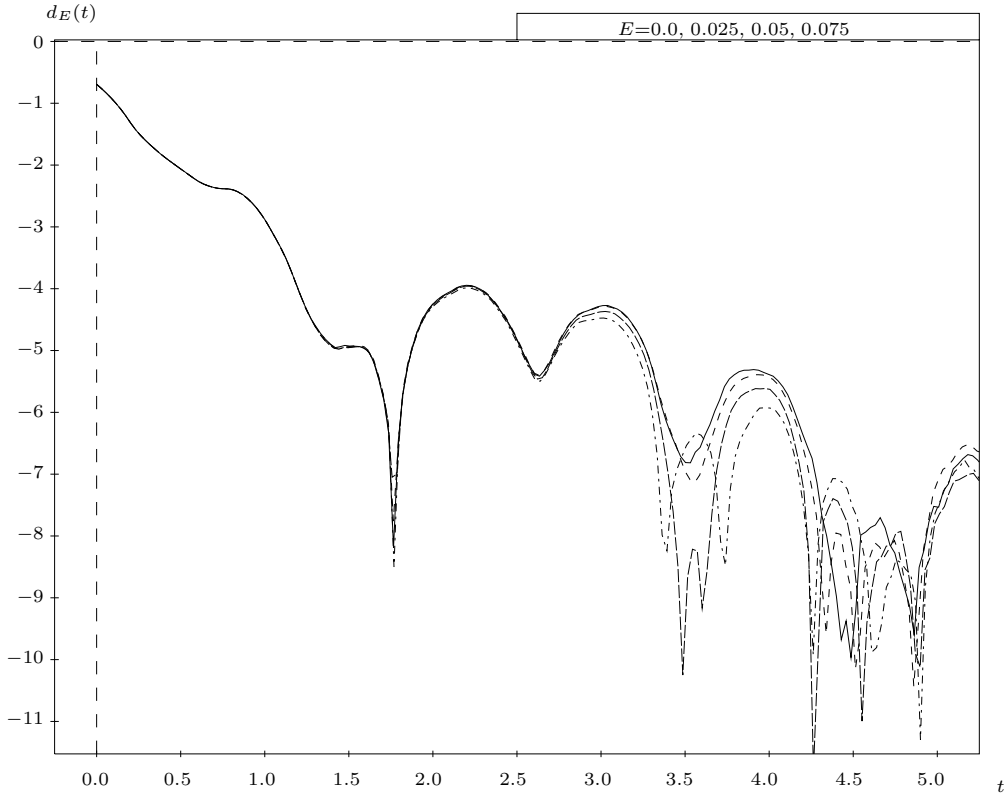
}{correla}{Logarithm of orrelation function with respect to the the 
Lebesgue measure for fields $E=0$ (solid line), $E=0.025$ (short dashed), 
$E=0.05$ (long dashed) and $E=0.075$ (dotted dashed).
}{fig10}0

\item{$\bullet$}An idea of this situation can come also from
\equ(prob). Indeed it is reasonable to think that $\Lambda_E(t,X)$ will
be given by a function that is smooth everywhere except on the
discontinuity generated by the collisions that happen before time
$t$. This implies that if $X$ is part of a grazing trajectory for a
given $E_0$ we can expect $\Lambda_{E}(t,X)$ to be non smooth as a
function of $E$ at $E=E_0$.  Considering that the set of grazing
trajectories varies smootly with $E$ and that \equ(prob) contains an
integral over the variable $X$ we do not expect any non-smoothness of
the current to take place before the set of grazing trajectories changes
its structure. A dimensional analysis gives the same result as for the
conjugation but we do not report it here because we do not believe that
the integrals in \equ(prob) converge. We show instead in
Fig. \equ(fig10) the behavior of $d_E(t)=\log(|D_E(t)|)$ with
$D_E(t)=\langle J(X)J(S_E(t,X))\rangle_E$ as a function of $t$ and
$E$.
We see that all these functions have a maxima for $t$ near 3 and near
4. The value at the maximum is a decreasing function of the field. In
general the correlation function appears to be decreasing in $E$ at
fixed $X$ except for the appearence of a new maximum around
$t=3.5$. We think that this new maximum can be connected with the
appearence of a new symbol in $D^3_E$ and so with the discontinuity in
the derivative of $\kappa(E)$ for small $E$. We leave this point as a
speculation because checking it is beyond the scope of this paper.

\newsec{Correlation function and diffusion coefficient}

In addition to $D_E(t)$ we studied the current-current correlation
function in the steady state:

$$C_E(t)=\langle J(X)J(S_E(t,X))\rangle_E-\langle J(X)\rangle_E^2$$

To compute this function we used a method similar to the one used for
computing the Kawasaki formula. After having generated an initial
point using the R250 generator we evolve it for 20 collisions with the
field $E=0$ to better decorrelate the initial choice. We then switch
on the field and let the system evolve for 50 collisions in such a way
that the final point can be considered as distributed according to the
stationary SRB distribution $\mu^+_E$. We then use such a point to compute a
trajectory segment of lenght 40 over wich we compute the
correlation. For small values of the field the correlation function
$C_E(t)$ appears very similar to the Kawasaki correlation
function. Considering that we already plotted $C_0(t)=D_0(t)$ in
Fig. \equ(fig_kk) we plot the logarithm of the correlation function
$c_E(t)=\log(|C_E(t)|)$ for $E=0.025$, $0.05$, $0.075$ and $0.1$.

\dimen2=25pt \advance\dimen2 by 126pt
\dimen3=212pt
\dimenfor{\dimen3}{166pt}
\eqfigfor{
\ins{75pt}{104pt}{$\hbox to60pt{\hfill ${\scriptstyle -0.86t-2.0}$\hfill}$}
\ins{192pt}{0pt}{$\hbox to20pt{\hfill ${\scriptstyle t}$\hfill}$}
\ins{12pt}{\dimen2}{$\hbox to25pt{\hfill ${\scriptstyle c_E(t)}$\hfill}$}
\ins{107pt}{144pt}{$\hbox to87pt{\hfill ${\scriptstyle E=0.025}$\hfill}$}
\ins{16pt}{3pt}{$\hbox to 19pt{\hfill${\scriptstyle 0.0}$\hfill}$}
\ins{35pt}{3pt}{$\hbox to 19pt{\hfill${\scriptstyle 0.5}$\hfill}$}
\ins{55pt}{3pt}{$\hbox to 19pt{\hfill${\scriptstyle 1.0}$\hfill}$}
\ins{75pt}{3pt}{$\hbox to 19pt{\hfill${\scriptstyle 1.5}$\hfill}$}
\ins{95pt}{3pt}{$\hbox to 19pt{\hfill${\scriptstyle 2.0}$\hfill}$}
\ins{115pt}{3pt}{$\hbox to 19pt{\hfill${\scriptstyle 2.5}$\hfill}$}
\ins{134pt}{3pt}{$\hbox to 19pt{\hfill${\scriptstyle 3.0}$\hfill}$}
\ins{154pt}{3pt}{$\hbox to 19pt{\hfill${\scriptstyle 3.5}$\hfill}$}
\ins{174pt}{3pt}{$\hbox to 19pt{\hfill${\scriptstyle 4.0}$\hfill}$}
\ins{0pt}{19pt}{$\hbox to 14pt{\hfill${\scriptstyle -9}$}$}
\ins{0pt}{33pt}{$\hbox to 14pt{\hfill${\scriptstyle -8}$}$}
\ins{0pt}{46pt}{$\hbox to 14pt{\hfill${\scriptstyle -7}$}$}
\ins{0pt}{60pt}{$\hbox to 14pt{\hfill${\scriptstyle -6}$}$}
\ins{0pt}{73pt}{$\hbox to 14pt{\hfill${\scriptstyle -5}$}$}
\ins{0pt}{87pt}{$\hbox to 14pt{\hfill${\scriptstyle -4}$}$}
\ins{0pt}{101pt}{$\hbox to 14pt{\hfill${\scriptstyle -3}$}$}
\ins{0pt}{114pt}{$\hbox to 14pt{\hfill${\scriptstyle -2}$}$}
\ins{0pt}{128pt}{$\hbox to 14pt{\hfill${\scriptstyle -1}$}$}
}{
\ins{75pt}{104pt}{$\hbox to60pt{\hfill ${\scriptstyle -0.94t-1.85}$\hfill}$}
\ins{192pt}{0pt}{$\hbox to20pt{\hfill ${\scriptstyle t}$\hfill}$}
\ins{12pt}{\dimen2}{$\hbox to25pt{\hfill ${\scriptstyle c_E(t)}$\hfill}$}
\ins{107pt}{144pt}{$\hbox to87pt{\hfill ${\scriptstyle E=0.05}$\hfill}$}
\ins{16pt}{3pt}{$\hbox to 19pt{\hfill${\scriptstyle 0.0}$\hfill}$}
\ins{35pt}{3pt}{$\hbox to 19pt{\hfill${\scriptstyle 0.5}$\hfill}$}
\ins{55pt}{3pt}{$\hbox to 19pt{\hfill${\scriptstyle 1.0}$\hfill}$}
\ins{75pt}{3pt}{$\hbox to 19pt{\hfill${\scriptstyle 1.5}$\hfill}$}
\ins{95pt}{3pt}{$\hbox to 19pt{\hfill${\scriptstyle 2.0}$\hfill}$}
\ins{115pt}{3pt}{$\hbox to 19pt{\hfill${\scriptstyle 2.5}$\hfill}$}
\ins{134pt}{3pt}{$\hbox to 19pt{\hfill${\scriptstyle 3.0}$\hfill}$}
\ins{154pt}{3pt}{$\hbox to 19pt{\hfill${\scriptstyle 3.5}$\hfill}$}
\ins{174pt}{3pt}{$\hbox to 19pt{\hfill${\scriptstyle 4.0}$\hfill}$}
\ins{0pt}{19pt}{$\hbox to 14pt{\hfill${\scriptstyle -9}$}$}
\ins{0pt}{33pt}{$\hbox to 14pt{\hfill${\scriptstyle -8}$}$}
\ins{0pt}{46pt}{$\hbox to 14pt{\hfill${\scriptstyle -7}$}$}
\ins{0pt}{60pt}{$\hbox to 14pt{\hfill${\scriptstyle -6}$}$}
\ins{0pt}{73pt}{$\hbox to 14pt{\hfill${\scriptstyle -5}$}$}
\ins{0pt}{87pt}{$\hbox to 14pt{\hfill${\scriptstyle -4}$}$}
\ins{0pt}{101pt}{$\hbox to 14pt{\hfill${\scriptstyle -3}$}$}
\ins{0pt}{114pt}{$\hbox to 14pt{\hfill${\scriptstyle -2}$}$}
\ins{0pt}{128pt}{$\hbox to 14pt{\hfill${\scriptstyle -1}$}$}
}{
\ins{75pt}{104pt}{$\hbox to60pt{\hfill ${\scriptstyle -1.06t-1.6}$\hfill}$}
\ins{192pt}{0pt}{$\hbox to20pt{\hfill ${\scriptstyle t}$\hfill}$}
\ins{12pt}{\dimen2}{$\hbox to25pt{\hfill ${\scriptstyle c_E(t)}$\hfill}$}
\ins{107pt}{144pt}{$\hbox to87pt{\hfill ${\scriptstyle E=0.075}$\hfill}$}
\ins{16pt}{3pt}{$\hbox to 19pt{\hfill${\scriptstyle 0.0}$\hfill}$}
\ins{35pt}{3pt}{$\hbox to 19pt{\hfill${\scriptstyle 0.5}$\hfill}$}
\ins{55pt}{3pt}{$\hbox to 19pt{\hfill${\scriptstyle 1.0}$\hfill}$}
\ins{75pt}{3pt}{$\hbox to 19pt{\hfill${\scriptstyle 1.5}$\hfill}$}
\ins{95pt}{3pt}{$\hbox to 19pt{\hfill${\scriptstyle 2.0}$\hfill}$}
\ins{115pt}{3pt}{$\hbox to 19pt{\hfill${\scriptstyle 2.5}$\hfill}$}
\ins{134pt}{3pt}{$\hbox to 19pt{\hfill${\scriptstyle 3.0}$\hfill}$}
\ins{154pt}{3pt}{$\hbox to 19pt{\hfill${\scriptstyle 3.5}$\hfill}$}
\ins{174pt}{3pt}{$\hbox to 19pt{\hfill${\scriptstyle 4.0}$\hfill}$}
\ins{0pt}{19pt}{$\hbox to 14pt{\hfill${\scriptstyle -9}$}$}
\ins{0pt}{33pt}{$\hbox to 14pt{\hfill${\scriptstyle -8}$}$}
\ins{0pt}{46pt}{$\hbox to 14pt{\hfill${\scriptstyle -7}$}$}
\ins{0pt}{60pt}{$\hbox to 14pt{\hfill${\scriptstyle -6}$}$}
\ins{0pt}{73pt}{$\hbox to 14pt{\hfill${\scriptstyle -5}$}$}
\ins{0pt}{87pt}{$\hbox to 14pt{\hfill${\scriptstyle -4}$}$}
\ins{0pt}{101pt}{$\hbox to 14pt{\hfill${\scriptstyle -3}$}$}
\ins{0pt}{114pt}{$\hbox to 14pt{\hfill${\scriptstyle -2}$}$}
\ins{0pt}{128pt}{$\hbox to 14pt{\hfill${\scriptstyle -1}$}$}
}{
\ins{75pt}{104pt}{$\hbox to60pt{\hfill ${\scriptstyle -1.12t-1.5}$\hfill}$}
\ins{192pt}{0pt}{$\hbox to20pt{\hfill ${\scriptstyle t}$\hfill}$}
\ins{12pt}{\dimen2}{$\hbox to25pt{\hfill ${\scriptstyle c_E(t)}$\hfill}$}
\ins{107pt}{144pt}{$\hbox to87pt{\hfill ${\scriptstyle E=0.1}$\hfill}$}
\ins{16pt}{3pt}{$\hbox to 19pt{\hfill${\scriptstyle 0.0}$\hfill}$}
\ins{35pt}{3pt}{$\hbox to 19pt{\hfill${\scriptstyle 0.5}$\hfill}$}
\ins{55pt}{3pt}{$\hbox to 19pt{\hfill${\scriptstyle 1.0}$\hfill}$}
\ins{75pt}{3pt}{$\hbox to 19pt{\hfill${\scriptstyle 1.5}$\hfill}$}
\ins{95pt}{3pt}{$\hbox to 19pt{\hfill${\scriptstyle 2.0}$\hfill}$}
\ins{115pt}{3pt}{$\hbox to 19pt{\hfill${\scriptstyle 2.5}$\hfill}$}
\ins{134pt}{3pt}{$\hbox to 19pt{\hfill${\scriptstyle 3.0}$\hfill}$}
\ins{154pt}{3pt}{$\hbox to 19pt{\hfill${\scriptstyle 3.5}$\hfill}$}
\ins{174pt}{3pt}{$\hbox to 19pt{\hfill${\scriptstyle 4.0}$\hfill}$}
\ins{0pt}{19pt}{$\hbox to 14pt{\hfill${\scriptstyle -9}$}$}
\ins{0pt}{33pt}{$\hbox to 14pt{\hfill${\scriptstyle -8}$}$}
\ins{0pt}{46pt}{$\hbox to 14pt{\hfill${\scriptstyle -7}$}$}
\ins{0pt}{60pt}{$\hbox to 14pt{\hfill${\scriptstyle -6}$}$}
\ins{0pt}{73pt}{$\hbox to 14pt{\hfill${\scriptstyle -5}$}$}
\ins{0pt}{87pt}{$\hbox to 14pt{\hfill${\scriptstyle -4}$}$}
\ins{0pt}{101pt}{$\hbox to 14pt{\hfill${\scriptstyle -3}$}$}
\ins{0pt}{114pt}{$\hbox to 14pt{\hfill${\scriptstyle -2}$}$}
\ins{0pt}{128pt}{$\hbox to 14pt{\hfill${\scriptstyle -1}$}$}
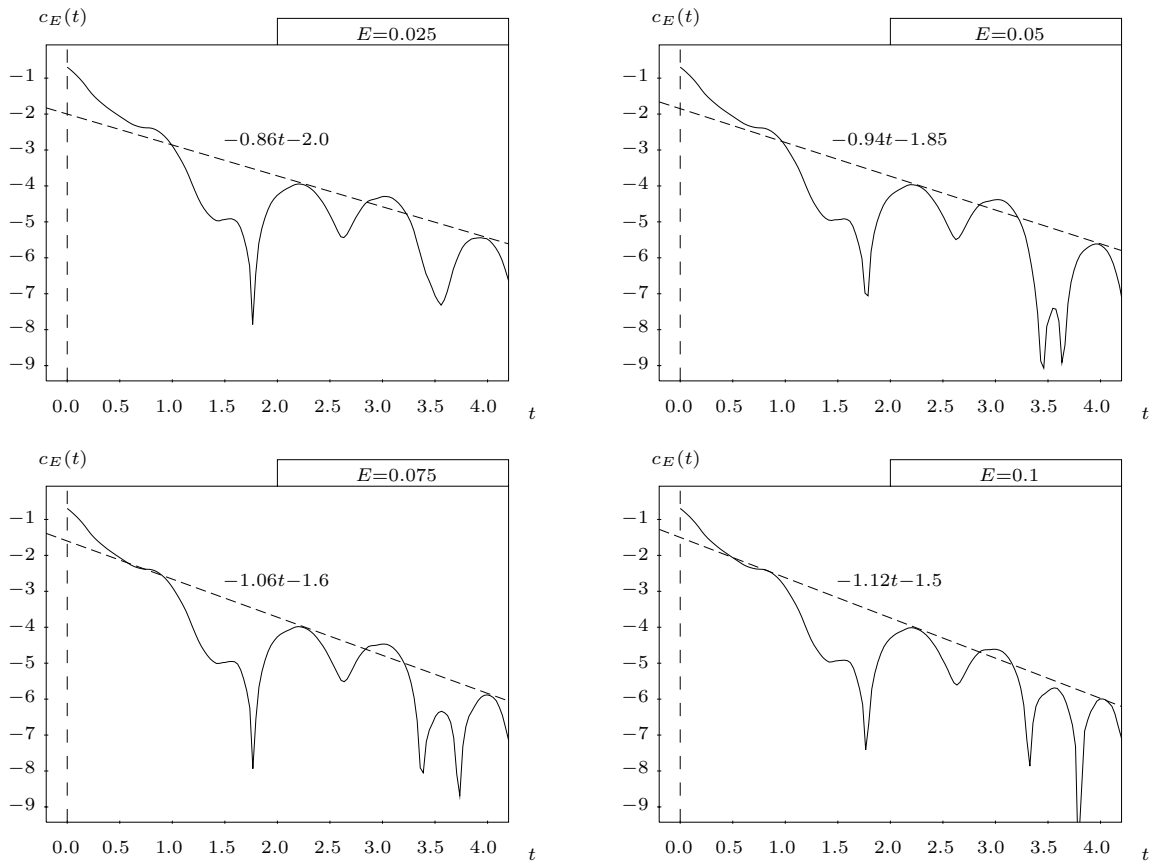
}{gk}{Estimation of the decay of the correlation function for $E=0.025$, 
$0.05$, $0.075$ and $0.1$ using a linear fit for the maxima of $c_E(t)$}{fig_gk}

One of the most interesting question about such correlation functions
is whether it decay exponentially. In the case $E=0$ it has been
proven in \cita{Y} and the proof can be extended for small (possibly
very small) value of the field \cita{C}. We use here a method
developed in \cita{GG} to analyze this question. In that paper the
exponential decay rate was obtained by dividing the set of maxima of
the function $c_0(t)$ into two groups and fitting them by a linear
law. Due to the fact that our simulations are shorter than the one
used in \cita{GG} (mainly for the reason that computing the collision
time for $E\not=0$ is much harder than for $E=0$ since the exact
solution contains transcendental functions) we see two maxima only for
one of the two groups used in \cita{GG}: the maximum near $t=2$ and the
one near $t=4$. While a fit for $E=0$ gives us a value consistent with
the one obtained by \cita{GG}\annotano{Note that our results are
reported in our adimensional time while in \cita{GG} they were
reported in unit of the mean free fly time} it is evident that the
slope of the fitting line becomes more negative (\ie a faster
exponential decay) when the field is switched on. A numerical value
can be deduced from the fit reported in Fig. \equ(fig_gk) but we do not
plot them separately because the error bar on each point would be too
large.

As already noted before the structure of $c_E(t)$ changes as $E$
changes. Moreover it seems to us (although with our data we cannot
really make a definitive statement) that one would still be able to
divide the maxima of the correlation function into groups each of
which well fitted by a linear law whose slope depends on
$E$\annotano{In \cita{GG} all the two groups of maxima where fitted by
linear laws with the same slope. It is unclear whether such a
property can be expected when $E\not=0$}. This behavior can be
explained if one assumes that the correlation function is given by a
sum of (quasi)-periodic functions modulated by decaying exponentials.
Such a behavior is well established in the case of a diamond billiard
by the numerical work in \cita{Ca}. The hypothesis formulated at the
end of the previous section will mean, in this setting, that the terms
in the above sum are linked with the symbolic dynamics of the
system. We hope to come back to this problem in a forthcoming work.

We computed the correlation function $C_E(t)$ also for values of the
field in the interval $(0.275,0.4)$. All the above discussion can be
applied to this case in the same way and we do not show any graphs
because no new information can be obtained from them.  The integral of
the correlation function $C_E(t)$ gives us the $xx$-element of the
diffusion matrix for the system, \ie $d(E)=\int_0^\infty C_E(t)dt$. We
conclude this paper reporting the value of this quantity obtained from
the above data.
%
\dimen2=25pt \advance\dimen2 by 269pt
\dimen3=398pt
\eqfig{\dimen3}{314pt}{
\ins{378pt}{0pt}{$\hbox to20pt{\hfill ${\scriptstyle E}$\hfill}$}
\ins{24pt}{\dimen2}{$\hbox to40pt{\hfill ${\scriptstyle \kappa(E)\backslash d(E)}$\hfill}$}
\ins{194pt}{287pt}{$\hbox to174pt{\hfill ${\scriptstyle \kappa(E)\;versus\;d(E)}$\hfill}$}
\ins{24pt}{3pt}{$\hbox to 42pt{\hfill${\scriptstyle 0.00}$\hfill}$}
\ins{66pt}{3pt}{$\hbox to 42pt{\hfill${\scriptstyle 0.05}$\hfill}$}
\ins{109pt}{3pt}{$\hbox to 42pt{\hfill${\scriptstyle 0.10}$\hfill}$}
\ins{151pt}{3pt}{$\hbox to 42pt{\hfill${\scriptstyle 0.15}$\hfill}$}
\ins{193pt}{3pt}{$\hbox to 42pt{\hfill${\scriptstyle 0.20}$\hfill}$}
\ins{235pt}{3pt}{$\hbox to 42pt{\hfill${\scriptstyle 0.25}$\hfill}$}
\ins{278pt}{3pt}{$\hbox to 42pt{\hfill${\scriptstyle 0.30}$\hfill}$}
\ins{320pt}{3pt}{$\hbox to 42pt{\hfill${\scriptstyle 0.35}$\hfill}$}
\ins{0pt}{35pt}{$\hbox to 26pt{\hfill${\scriptstyle 0.170}$}$}
\ins{0pt}{74pt}{$\hbox to 26pt{\hfill${\scriptstyle 0.172}$}$}
\ins{0pt}{114pt}{$\hbox to 26pt{\hfill${\scriptstyle 0.174}$}$}
\ins{0pt}{154pt}{$\hbox to 26pt{\hfill${\scriptstyle 0.176}$}$}
\ins{0pt}{193pt}{$\hbox to 26pt{\hfill${\scriptstyle 0.178}$}$}
\ins{0pt}{233pt}{$\hbox to 26pt{\hfill${\scriptstyle 0.180}$}$}
\ins{0pt}{273pt}{$\hbox to 26pt{\hfill${\scriptstyle 0.182}$}$}
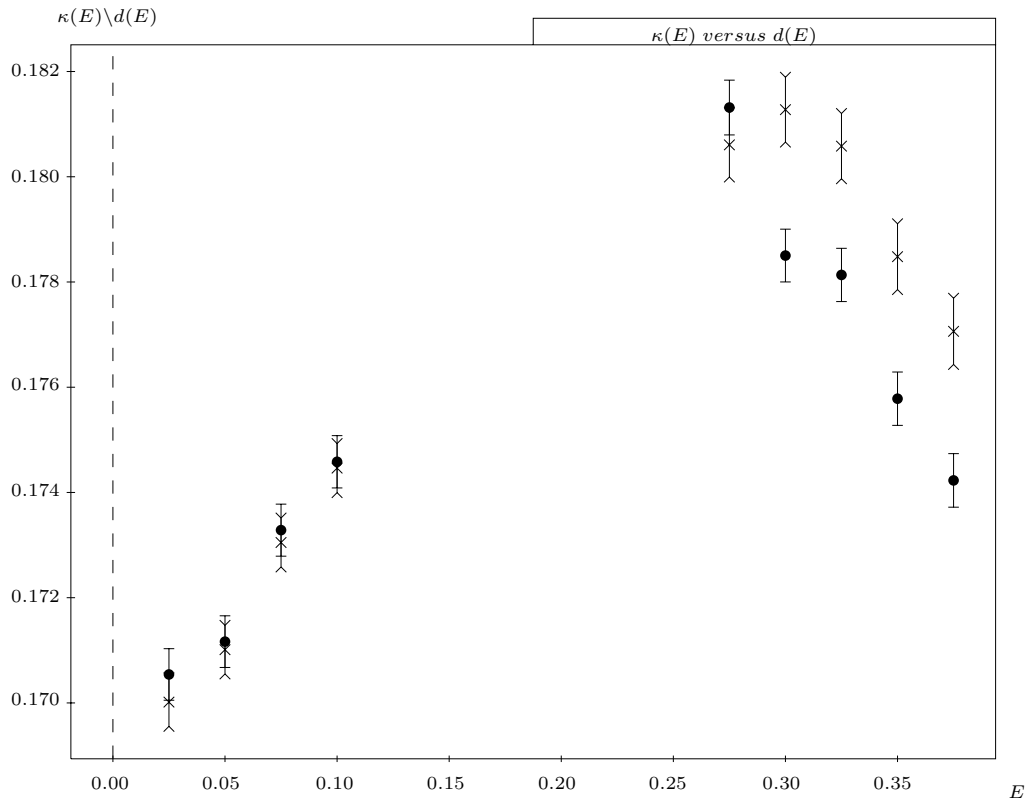
}{kk}{$d(E)$ versus $\kappa(E)$ showing good agreement near $E=0$ but clear
non equality at $E\not=0$}{fig_kk}0
As for the correlation function we have the data for $E\in(0.025,0.1)$
and $E\in(0.275,0.375)$. In Fig. \equ(fig_kk) the value of the
diffusion coefficient is plotted together with the value of 
$\kappa(E)$ obtained from the Kawasaki formula. It is clear that for
small fields the two formulas give almost the same value. 
Such an equality for $E\not=0$ would support the 
possibility of the validity of a Green-Kubo relation out of
equilibrium\cita{GG}\cita{ES}. Our results show that the
Kawasaki formula and the Green-Kubo relation give different results
for $E\not=0$.

\newsec{Conclusions}

In this paper we presented results of numerical simulations as well as 
analytical evidence for a non smooth behavior of
the current as a function of the field in a thermostatted single particle 
model of electrical conduction. 
We mainly studied the invariant measure $\nu^+_E$ for the
discrete time dynamical system obtained by considering as timing
events the collisions of the particle with the obstacles. Our
analysis is based on (at the present nonrigorous) construction of 
a conjugation map or
rather a ``partial conjugation map'' between the
dynamics at two different values of the field $E$. 
This construction strongly relies on the expectation that it is possible to
 analyze the dynamics  using symbols directly connected to
the discontinuities of the collision map. The picture can probably
be substantiated by constructing in a explicit way a Markov partition
for the dynamical system. 

Our analysis implies that the average with respect to $\nu^+_E$
of any smooth function on $\TT$ or, more generally, the average of
every smooth function on phase space with respect to $\mu^+_E$ is a not
twice differentiable with respect to the electric field $E$. Moreover one
can conjecture that the first derivative exists in a weak sense but is
everywhere discontinuous. Such a behavior has already been observed in
\cita{KD} for a very simple model.

While our analysis is not rigorous it  has the
advantage of being, at least in principle, contructive. 
The scheme for the construction of a ``partial conjugation'' or better
of a Markov partition can lead to a rigorous proof of our assertions
together with the possibility of conducting computer assisted
experiments on the billiards.

As noted earlier the recent formalism 
developed by Ruelle in \cita{Ru}
suggests the presence of the same kind of non smoothness but we do not
see a possibility of using it in a convincing way to argue in one
direction or the other. It seems to us that a form of conjugation
would be in any case necessary to show the convergence of eq.\equ(dkaw0).

We also analyze the behavior of the velocity-velocity correlation
function and of the diffusion coefficient. Applying the methods
developed in \cita{GG} we argue that the correlation functions decay
exponentially also when the field is non zero. Moreover the rate of
decay appears to increase with the field, at least for small values of $E$. 
The detailed structure of the
correlation function undergoes interesting changes and we try to
correlate these changes with the property of the
invariant measure discussed above. 
Due to the numerical difficulties of computing
correlation functions our discussion remains speculative.

Finally we computed the diffusion coefficient. For $E=0$ the
diffusion coefficient gives the linear conductivity. 
It has been argued that this relation could be
valid also for a field different from 0, see \cita{ES}. This would
imply that the integral of the correlation function computed with respect
to the Lebesgue measure, $d_E(t)$, and with respect to the SRB measure
$\mu_E^+$, should be equal. Our results show that this is not the case
so that no direct extention of a Green-Kubo formula to non-equilibrium
is possible (see \cita{G} for an interesting proposal of what
exending Green-Kubo out of equilibrium means). Results similar to ours were
already present in \cita{ES}. Considering that no good reason has ever
been given for the above equality to hold we consider this result as
natural. 

\0{\bf Acknowledgment} We are greatly
indebted to G. Gallavotti, N. Chernov, L.-S. Young, D. Ruelle and
C. Dettman for many enlightening discussion. Research
supported in part by NSF Grant DMR-9813268. 
D.D. aknowledges support from the Francqui Foundation. 
We aknowledge support and ospitatlity from the IHES where this work 
was finished.

\rife{Ca}{ACG}{R. Artuso, G. Casati, I. Guarneri: Geometric scaling of
correlation decay in chaotic billiards, {\it Phys. Rev. E} {\bf 51}
(1995) R3807-R3810}
\rife{BDGL}{BDGL}{F. Bonetto, D. Daems, J.L. Lebowitz, V. Ricci: 
Properties of Stationary Nonequilibrium States in the Thermostatted
Periodic Lorentz Gas II:  The many point particles system, in preparation}
\rife{BDLR}{BDLR}{F. Bonetto, D. Daems, P.L. Garrido, J.L. Lebowitz:
Properties of Stationary Nonequilibrium States in the Thermostatted
Periodic Lorentz Gas III:  The many colliding particles system, 
in preparation} 
\rife{BGG}{BGG}{F. Bonetto, G. Gallavotti, P.L. Garrido: Chaotic
Hypothesis: an experimental test {\it Physica D} {\bf 105} (1997), 226-252}
\rife{BKL}{BKL}{F. Bonetto, A. Kupiainen, J.L. Lebowitz:  
Perturbation theory for coupled Arnold cat maps: absolute
continuity of marginal distribution, in preparation}
\rife{L}{BL}{P.G. Bergaman, J.L. Lebowitz: {\it Phys. Rev.} {\bf 99} (1955) 2}
\rife{C}{C}{N. Chernov: Sinai biliards under small external forces, 
preprint, available at \hfill\penalty-10000{\tt http://www.math.uab.edu/chernov/pubs.html}}
\rife{CELS}{CELS}{N. Chernov, G. Eyink, J.E. Lebowitz, Ya.G. Sinai: 
Steady state electric conductivity in the periodic Lorentz gas, 
{\it Commun. in Math.  Phys.} {\bf 154} (1993), 569--601.}
\rife{EM}{EM}{D.J. Evans, G.P. Morris: {\it Statistical Mechanics of nonequilibrium 
fluids}, Accademic Press, San Diego (1990)}
\rife{EPR1}{EPR1}{J. P. Eckmann, C.-A. Pillet and L. Rey-Bellet:
Entropy production in nonlinear, thermally driven
Hamiltonian systems. {\it J. Statist. Phys.} {\bf 95} (1999) 305--331.}
\rife{EPR2}{EPR2}{J. P. Eckmann, C.-A. Pillet and L. Rey-Bellet:
Non-equilibrium statistical mechanics of anharmonic
chains coupled to two heat baths at different temperatures. 
{\it Comm. Math. Phys.} {\bf 201} (1999) 657--697.}
\rife{Po}{ER}{J.P. Eckmann, D. Ruelle: Ergodic Theory of Chaos and
Strange Attractors, {\it Rev. Mod. Phys.} {\bf 57} (1985) 617--656.}
\rife{ES}{ES}{D. Searles,D.J. Evans: The Fluctuation Theorem and Green-Kubo 
relations, to appear in JCP}
\rife{G}{G}{G. Gallavotti: Extension of Onsager's reciprocity relations 
to large fields and the chaotic hypothesis {\it Phys. Rev. Lett} {\bf 77} (1996) 
4334-4337}
\rife{GC}{GC}{G. Gallavotti, E.G.D. Cohen: Dynamical ensamble in stationary 
states, {\it J. Stat. Phys.} {\bf 80} (1995) 931}
\rife{GG}{GG}{G. Gallavotti, P.L. Garrido: Billiard
correlation-functions, {\it J. Stat. Phys.} {\bf 76} (1994) 549-585}
\rife{H}{H}{W.G. Hoover: {\it Computational Statistical Mechanics}, Elsevier (1991)}
\rife{LL}{JLl}{M. Jiang, R. de la LLave: Smooth dependence of
thermodynamic limits of SRB-measures mp$\underline{\ }$arc/99-122}
\rife{KD}{KD}{R. Klages, J.R. Dorfman: Simple Maps with Fractal
Diffusion Coefficients, {\it Phys. Rew. Lett.} {\bf 74} (1995) 387-390}
\rife{R}{LNRM}{J. Lloyd, M. Niemeyer, L. Rondoni, G.P. Morris: The
non-equilibrium Lorentz gas, {\it Chaos} {\bf 5} (1995) 536-551}
\rife{LS}{LS}{J.L. Lebowitz, H. Spohn: A Gallavotti-Cohen type symmetry in the 
large deviation functional for Stochastic Dynamics, {\it J. Stat. Phys.} 
{\bf 95} (1999) 333}
\rife{MH}{MH}{B. Moran, W.G. Hoover: Diffusion in the periodic Lorentz
billiard {\it J. Stat. Phys.} {\bf 48}(1987) 709}
\rife{NR}{NR}{W.H. Press, B.P. Flannery, S.A. Teukolsky,
W.T. Wetterling {\it Numerical recipes in C}, Cambridge University
Press (1993)}
\rife{Ru}{R}{D. Ruelle: Differentiation of SRB states
{\it Commun. Math. Phys.} {\bf 187} (1997) 227-241 and 
Nonequilibrium statistical mechanics near equilibrium: computing 
higher-order terms {\it Nonlinearity} {\bf 11} (1998) 5-18}
\rife{W}{W}{M. Wojtkowski: W-flows on Weyl manifolds and Gaussian 
thermostat, preprint, available at \hfill\penalty-10000
{\tt http://www.ma.utexas.edu/mp$\underline{\;}$arc} \#00-51}
\rife{Y}{Y}{L.-S. Young: Statistical properties of systems with
some hyperbolicity including certain billiards, {\it Annals of Math.}
{\bf 147} (1998) 585-650}

\biblio


\bye